\documentclass[11pt]{article}

\usepackage{amsmath}
\usepackage{amsfonts}
\usepackage{amssymb}
\usepackage{amsthm}
\usepackage{graphicx}
\usepackage{cite}
\usepackage{empheq}
\usepackage{caption}
\usepackage{subcaption}

\newtheorem{theorem}{Theorem}

\newtheorem{definition}{Definition}
\newtheorem{remark}{Remark}

\title{Higher-order in time ``quasi-unconditionally stable'' ADI solvers for the compressible Navier-Stokes equations\\ in 2D and 3D curvilinear domains}

\author{Oscar P. Bruno and Max Cubillos}

\begin{document}

\date{}

\maketitle

\begin{abstract}
This paper introduces alternating-direction implicit (ADI) solvers of
higher order of time-accuracy (orders two to six) for the compressible
Navier-Stokes equations in two- and three-dimensional curvilinear
domains. The higher-order accuracy in time results from 1)~An
application of the backward differentiation formulae time-stepping
algorithm (BDF) in conjunction with 2)~A BDF-like extrapolation
technique for certain components of the nonlinear terms (which makes
use of nonlinear solves unnecessary), as well as 3)~A novel
application of the Douglas-Gunn splitting (which greatly facilitates
handling of boundary conditions while preserving higher-order accuracy
in time).  As suggested by our theoretical analysis of the algorithms
for a variety of special cases, an extensive set of numerical
experiments clearly indicate that all of the BDF-based ADI algorithms
proposed in this paper are ``quasi-unconditionally stable'' in the
following sense: {\em each algorithm is stable for all couples
  $(h,\Delta t)$ of spatial and temporal mesh sizes in a
  problem-dependent rectangular neighborhood of the form
  $(0,M_h)\times (0,M_t)$}.  In other words, for each {\em fixed}
value of $\Delta t$ below a certain threshold, the Navier-Stokes
solvers presented in this paper are stable for arbitrarily small
spatial mesh-sizes. The second order formulation has further been
rigorously shown to be unconditionally stable for linear hyperbolic
and parabolic equations in two-dimensional space.  Although implicit
ADI solvers for the Navier-Stokes equations with nominal second-order
of temporal accuracy have been proposed in the past, the algorithms
presented in this paper are the first ADI-based Navier-Stokes solvers
for which second order or better accuracy has been verified in
practice under non-trivial (non-periodic) boundary conditions.
\end{abstract}

\section{Introduction} \label{sec:Introduction}



The direct numerical simulation of fluid flow at high Reynolds numbers
presents a number of significant
challenges~\cite{garnier_large_2009}---including the presence of
structures such as boundary layers, eddies, vortices and turbulence,
whose accurate spatial discretization requires use of fine spatial
meshes. Simulation of such flows by means of explicit solvers is
difficult, even on massively parallel super computers, in view of the
severe restrictions on time-steps required for stability: the
time-step must scale like the square of the spatial
mesh-size. Classical implicit solvers do not suffer from such
time-step restrictions but they do require solution of large systems
of equations at each time step, and they can therefore be extremely
expensive as well.  The celebrated Beam and Warming
method~\cite{beam_implicit_1976,beam_implicit_1978} provides one of
the most attractive alternatives to explicit and classical implicit
algorithms. Based on the the Alternating Direction Implicit
method~\cite{peaceman_numerical_1955} (ADI), the Beam and Warming
scheme enables stable solution of the compressible Navier-Stokes
equations without recourse to either nonlinear iterative solvers or
solution of large linear systems at each time-step.

Significant
efforts~\cite{visbal_high-order-accurate_1999,ekaterinaris_implicit_1999,rizzetta_high-order_2008,gordnier_high_2009,uzun_simulation_2009,kawai_large-eddy_2010}
have centered around the ideas first put forth in the celebrated
papers~\cite{beam_implicit_1976,beam_implicit_1978}, focusing, in
particular, on enhancing stability and restoring the (nominal)
second-order of accuracy inherent in the original derivation of the
method. (The discussion in Section~\ref{sec:BCDiscussion} suggests that,
indeed, the overall accuracy of a time-marching ADI scheme based on
second-order time-stepping may drop to first-order of accuracy in time
if boundary conditions at intermediate-times are not enforced with the
correct accuracy order; see
also~\cite{briley_structure_1980,leveque_intermediate_1985}.)  The
aforementioned modifications of the
algorithms~\cite{beam_implicit_1976,beam_implicit_1978} incorporate
various kinds of Newton-like subiterations to reduce the errors
arising from the approximation of the nonlinear terms while
maintaining stability. Unfortunately, however, no numerical examples
have been provided that demonstrate second-order time-accuracy for the
modified algorithms---even though in all such cases nominally
second-order time-stepping schemes are used. Still, as demonstrated by
the numerical results presented in this paper (see
e.g. Figures~\ref{fig:ManufConv3DWavy},~\ref{fig:AnnulusConvBDF2}
and~\ref{fig:AnnulusConvBDF3} and associated discussion), high-order
time accuracy may be highly advantageous in the treatment of long-time
simulations or highly-inhomogeneous flows---for which the temporal
dispersion inherent in low-order approaches would make it necessary to
use inordinately small time-steps.

The present paper (Part~I in a two-part contribution) introduces ADI
solvers of higher orders of time-accuracy (orders $s=2$ to 6) for the
compressible Navier-Stokes equations in two- and three-dimensional
curvilinear domains; for definiteness spectral Chebyshev and Fourier
collocation spatial discretizations are used throughout this
paper. The new ADI algorithms successfully address the difficulties
discussed above: (i)~They (provably) enjoy high orders of
time-accuracy (orders two to six) {\em even in presence of general
  (and, in particular, non-periodic) boundary conditions}; (ii)~They
do not require use of iterative nonlinear solvers for accuracy or
stability, and they rely, instead, on a BDF-like extrapolation
technique~\cite{bruno_higher-order_2014} for certain components of the
nonlinear terms; and, as established in
Part~II~\cite{bruno_quasi-unconditional_2015}, (iii)~They possess
favorable stability properties, with {\em rigorous
  unconditional-stability proofs for constant coefficient hyperbolic
  and parabolic equations} for $s=2$, and demonstrating in practice
{\em quasi-unconditional stability for $2\leq s\leq 6$}
(Definition~\ref{def:QuasiUnconditional} in Section~\ref{sec:BDFQuasiUnconditional}).

The ADI split and extrapolation techniques mentioned above give rise
to certain two-point boundary-values problems for second-order systems
of ODEs with {\em variable} coefficients. The specific methodologies
utilized to solve such ODE systems depends significantly on the method
used for spatial approximation. If a finite-difference approximation
is used, then the solution of the ODE systems under consideration
entails inversion of a linear system with a banded system matrix,
which can be treated efficiently by means of sparse LU
decompositions~\cite{golub_matrix_2012}. If spectral methods are used,
as in this paper, in turn, then the ODE solution requires inversion of
a full non-sparse linear system. In order to tackle such problems
efficiently we resort to use of the GMRES
algorithm~\cite{saad_gmres:_1986} in conjunction with a
finite-difference preconditioner, as detailed in
Section~\ref{sec:DiscreteSpatialDetails}. The resulting overall Navier-Stokes solvers are
applicable to curvilinear coordinate systems in general domains; an
accuracy-order-preserving spectral filter is used in our scheme to
ensure stability.

Although quite fast when compared to other implicit algorithms, the
implicit solvers introduced in this paper are more expensive per
time-step than a similarly discretized explicit solver. Nevertheless,
implicit solvers of high-order time accuracy such as the ones
introduced in what follows may play crucial roles in the treatment of
long-time simulations, highly-inhomogeneous flows, complex physical
boundaries as well as steep boundary layers. For such problems the
requirements of fine spatial discretization meshes---which may be
needed for flow and/or geometry resolution---could necessitate use of
very small time-steps, in view of the restrictive CFL constraints
inherent in explicit solvers for Navier-Stokes problems, and thus,
possibly, to inordinately large computing times.  Extensions of these
algorithms to the multi-domain overlapping-patch
context~\cite{brown_overture:_1999} as well as other types of spatial
discretizations, including
finite-differences~\cite{leveque_finite_2007} and Fourier
Continuation~\cite{bruno_high-order_2010,albin_spectral_2011}, are
currently in progress and will be presented elsewhere. In particular,
the present curvilinear domain algorithms are ideally suited as
single-domain implicit components of general multi-domain
implicit/explicit solvers~\cite{cubillos_general-domain_2015}.

This paper is organized as follows: after notations are introduced in
Section~\ref{sec:Preliminaries}, Section~\ref{sec:BDF-ADI} presents the proposed
BDF-ADI methodology in full
detail. Section~\ref{sec:StabilityDiscussion} discusses the stability
properties of the proposed methods, with reference to the stability
proofs and numerical demonstrations put forth in
Part~II. Section~\ref{sec:NumericalImplementation} then presents
details of our algorithmic implementations, and
Section~\ref{sec:NumericalResults} demonstrates the stability and
accuracy of the proposed algorithms by means of a variety of numerical
results. Concluding remarks, finally, are presented in
Section~\ref{sec:Conclusions}.


\section{Preliminaries\label{sec:Preliminaries}}

We consider the compressible Navier-Stokes equations for the velocity
$\mathbf{u}$, temperature $T$ and density $\rho$ in a perfect gas in a
$d$-dimensional domain $\Omega \subset \mathbb{R}^d$ ($d = 2$ or $d =
3$).  For definiteness we assume the pressure $p$, density $\rho$,
temperature $T$ are related by the ideal gas law $p = \rho R T$ with
gas constant $R$. The heat flux $q$ and the temperature gradient
$\nabla T$, in turn, are assumed to satisfy the isotropic Fourier law
$q = -\kappa \, \nabla T$ for a certain (temperature-dependent)
thermal conductivity constant $\kappa$. Using characteristic values
$L_0$, $u_0$, $\rho_0$, $T_0$, $\mu_0$ and $\kappa_0$ for length,
velocity, density, temperature, viscosity and heat conductivity,
respectively, the Navier-Stokes equations under consideration can be
expressed in the non-dimensional form~\cite{white_viscous_2006}
\begin{subequations} \label{eq:NavierStokes}
\begin{eqnarray}
	\rho_t + \nabla \cdot (\rho \mathbf{u}) &=& 0 \label{eq:NavierStokesContinuity} \\
	\mathbf{u}_t + \mathbf{u} \cdot \nabla \mathbf{u} + \frac{1}{\gamma\mathrm{Ma}^2} \frac{1}{\rho} \nabla (\rho T) &=& \frac{1}{\mathrm{Re}} \frac{1}{\rho} \nabla \cdot \boldsymbol{\sigma} \label{eq:NavierStokesMomentum} \\
	T_t + \mathbf{u} \cdot \nabla T + (\gamma - 1) T \nabla \cdot \mathbf{u} &=& \frac{\gamma}{\mathrm{Re}\mathrm{Pr}} \frac{1}{\rho} \nabla \cdot (\kappa \nabla T) + \frac{\gamma(\gamma - 1)\mathrm{Ma}^2}{\mathrm{Re}} \frac{1}{\rho} \Phi \label{eq:NavierStokesEnergy} 
\end{eqnarray} 
\end{subequations}
($\mathbf{u} = (u_i)$, $\boldsymbol{\sigma} = (\sigma_{ij})$,
$i,j=1,\dots,d$) where $\gamma = c_p / c_v$ is the ratio of specific
heats, $\mathrm{Re} = \rho_0 u_0 L_0 / \mu_0$ is the Reynolds number
and $\mathrm{Ma} = u_0 / \sqrt{\gamma R T_0}$ is the Mach number (with
gas constant $R = c_p - c_v$), and where $\mathrm{Pr} = \mu_0 c_p /
\kappa_0$ is the Prandtl number. The non-dimensional primitive
variables in these equations are the velocity vector $\mathbf{u}$, the
density $\rho$ and the temperature $T$; the quantities
$\boldsymbol{\sigma}$ and $\Phi$, in turn, denote the Newtonian
deviatoric stress tensor and the viscous dissipation function,
respectively: letting $I$ denote the identity tensor, we have
$$ \boldsymbol{\sigma} = \mu \left( \nabla \mathbf{u} + \nabla \mathbf{u}^{\mathrm T} - \frac{2}{3}(\nabla \cdot \mathbf{u}) \mathbf{I} \right)\quad\mbox{and}\quad \Phi = \sum_{ij}\sigma_{ij} \partial_{x_i}u_j .$$
For definiteness we assume $\mu$ and $\kappa$ are functions of
temperature alone---as they are, for example, under Sutherland's
law~\cite[pp. 28--30]{white_viscous_2006}
\begin{equation} \label{eq:Sutherland}
  \kappa = \frac{1+S_\kappa}{T+S_\kappa}T^{3/2} \quad\mbox{and}\quad  \mu = \frac{1+S_\mu}{T+S_\mu}T^{3/2},
\end{equation}
where $S_\kappa$ and $S_\mu$ are the non-dimensionalized Sutherland constants.

Letting $Q = (\mathbf{u}^{\mathrm T},T,\rho)^{\mathrm T} $ denote the
full $(d+2)$-dimensional solution vector,
the equations~\eqref{eq:NavierStokes} can be expressed in the form
\begin{equation}\label{eq:GeneralIVP}
	Q_t = \mathcal{P}(Q,t)\quad,\quad x \in \Omega\quad ,\quad t\geq 0
\end{equation}
where $\mathcal{P}$ is a vector-valued nonlinear differential
operator.  Note the $t$ dependence in the operator $\mathcal P$ which
allows for the presence of time-dependent source terms. The system is
completed by means of the relevant boundary conditions for a given
configuration; see e.g.~\cite[Sec. 1-4]{white_viscous_2006}.

\begin{remark} \label{rem:InflowBC} For notational simplicity our
  description of the BDF-ADI algorithms assumes that no-slip
  boundary conditions of the form
\begin{equation} \label{eq:BoundaryConditions-no-sl}
\left .
\begin{pmatrix}
  \mathbf{u} \\  T
\end{pmatrix}
\right|_{\partial D}
=
\begin{pmatrix}
	g^\mathbf{u} \\ g^T
\end{pmatrix}
\end{equation}
are prescribed, where $g^\mathbf{u}$ and $g^T$ are given functions
defined on $\partial D$. Certainly, other relevant types of boundary
conditions can be incorporated within the proposed
framework---Section~\ref{sec:NumericalResults} includes an example of unsteady boundary layer flow that incorporates
no-slip boundary conditions at a rough boundary as well as inflow and
absorbing boundary conditions.
\end{remark}

\section{BDF-ADI methodology} \label{sec:BDF-ADI}

This section introduces the proposed BDF-ADI approach. Throughout this
section derivations and formulae are given for problems in $d=3$
spatial dimensions but, in all cases, the treatment can be applied
easily to obtain the corresponding two-dimensional counterparts. The
numerical examples in Section~\ref{sec:NumericalResults}, in
particular, include applications to problems in both two- and
three-dimensional space.

\subsection{Quasilinear-like Cartesian formulation \label{sec:QuasiCartesian}}

The derivation of the proposed BDF-ADI method relies on a certain
quasilinear-like formulation of the Navier-Stokes equations. To
motivate the introduction of this concept we note that provided $\mu$
and $\kappa$ are constant and the viscous dissipation function $\Phi$
is neglected, the Navier-Stokes equations under consideration may be
expressed in the quasilinear form
\begin{align} \label{eq:NavierStokesCartesian3D}
	Q_t & + M^{x,1}(Q) \frac{\partial}{\partial x} Q + M^{y,1}(Q) \frac{\partial}{\partial y} + M^{z,1}(Q) \frac{\partial}{\partial z} Q \nonumber \\
	    & + M^{x,2}(Q) \frac{\partial^2}{\partial x^2} Q + M^{y,2}(Q) \frac{\partial^2}{\partial y^2} Q + M^{z,2}(Q) \frac{\partial^2}{\partial z^2} Q \nonumber \\
	    & + M^{xy}(Q) \frac{\partial^2}{\partial x \partial y} Q + M^{xz}(Q) \frac{\partial^2}{\partial x \partial z} Q + M^{yz}(Q) \frac{\partial^2}{\partial y \partial z} Q + M^0(Q) Q \nonumber \\
	    & = 0,
\end{align}
where the various $M$ matrices ($M^{x,1}$, $M^{x,2}$ etc.) are
matrix-valued functions of $Q$. Such a formulation can be made to
stand even if $\Phi$ is not neglected and/or $\mu$ and $\kappa$ are
non-constant, provided the $M$ matrices are allowed to incorporate
certain derivative terms. For example, terms such as $u_x^2$ which arise in the
dissipation function $\Phi$ can be incorporated in such a formulation
by including one $u_x$ term in the matrix $M^{x,1}$ and the second
$u_x$ term in the vector $\partial_x Q$.  Similarly, expanding the
product $ \mu(T)_x u_y$ (that arises in in the term $\nabla \cdot
\boldsymbol{\sigma}$) by means of the chain rule,
\begin{align}\label{eq:NonlinearProduct} 
  \mu(T)_x u_y &= \mu'(T) T_x u_y\nonumber \\
              &= \left( \frac{1}{2}\mu'(T) T_x \right) u_y + \left( \frac{1}{2}\mu'(T) u_y \right) T_x,
\end{align}
the two quantities in parentheses can be included in the matrices
$M^{y,1}$ and $M^{x,1}$ respectively. (We note that while the
expression~\eqref{eq:NonlinearProduct} treats the factors in the product $T_x
u_y$ in an symmetric manner, other alternatives may be useful as
well.) The $M$ matrices resulting from this approach are presented in
Appendix~\ref{app:NSMatrices}.

Upon temporal discretization the proposed algorithms treat implicitly
all spatial derivatives in
equations~\eqref{eq:NavierStokesCartesian3D} not included in the $M$
matrices, and they approximate the corresponding $M$ coefficients by
means of certain explicit time extrapolations.  Full algorithmic
details in these regards, allowing for possible use of curvilinear
coordinates, are presented in sections~\ref{sec:QuasiCurvilinear}
through~\ref{sec:BCDiscussion}.

\subsection{Quasilinear-like curvilinear formulation} \label{sec:QuasiCurvilinear}

Let $x=x(\xi,\eta,\zeta)$, $y=y(\xi,\eta,\zeta)$,
$z=z(\xi,\eta,\zeta)$ define a smooth invertible mapping from the the
$(\xi,\eta,\zeta)$ computational domain (which we take to be the cube
$D = [\ell_1,\ell_2]^3$ for some real numbers $\ell_1$ and $\ell_2$)
to the physical domain $\Omega\subset \mathbb{R}^3$.  With minor
notational abuses, we will denote the corresponding inverse mapping by
$\xi=\xi(x,y,z)$, $\eta=\eta(x,y,z)$, $\zeta=\zeta(x,y,z)$, and we
will write $Q(\xi,\eta,\zeta) =
Q(x(\xi,\eta,\zeta),y(\xi,\eta,\zeta),z(\xi,\eta,\zeta))$.  A
formulation in terms of the variables $(\xi,\eta,\zeta)$ is obtained
by invoking the chain rule: the derivatives with respect to $x$, $y$,
and $z$ in equation~\eqref{eq:NavierStokesCartesian3D} are expressed in
terms of derivatives with respect to $\xi$, $\eta$, and $\zeta$ of the
unknown $Q$ and certain ``metric terms'' given by derivatives of
$\xi$, $\eta$, and $\zeta$ with respect to the Cartesian
variables. Using these variables in~\eqref{eq:NavierStokesCartesian3D}
and collecting terms we obtain an equation for
$Q=Q(\xi,\eta,\zeta,t)$:
\begin{align} \label{eq:NavierStokesCurvilinear3D}
  Q_t & + M^{\xi,1}(Q) \frac{\partial}{\partial \xi} Q + M^{\eta,1}(Q) \frac{\partial}{\partial \eta} + M^{\zeta,1}(Q) \frac{\partial}{\partial \zeta} Q \nonumber \\
  & + M^{\xi,2}(Q) \frac{\partial^2}{\partial \xi^2} Q + M^{\eta,2}(Q) \frac{\partial^2}{\partial \eta^2} Q + M^{\zeta,2}(Q) \frac{\partial^2}{\partial \zeta^2} Q \nonumber \\
  & + M^{\xi\eta}(Q) \frac{\partial^2}{\partial \xi \partial \eta} Q + M^{\xi\zeta}(Q) \frac{\partial^2}{\partial \xi \partial \zeta} Q + M^{\eta\zeta}(Q) \frac{\partial^2}{\partial \eta \partial \zeta} Q + M^0(Q) Q \nonumber \\
  & = 0
\end{align}
for $(\xi,\eta,\zeta) \in D$, where the matrices $M^{\xi,1}(Q)$
(which, per the discussion in Section~\ref{sec:QuasiCartesian},
generally contain derivatives of $Q$ with regards to
$(\xi,\eta,\zeta)$) can be obtained by incorporating the various
metric terms in the corresponding Cartesian matrices; see
e.g.~\cite{hoffmann_computational_2000}. Explicit expressions for the
$M$ matrices in equation~\eqref{eq:NavierStokesCurvilinear3D} are
presented in Appendix~\ref{app:NSMatrices}.

\subsection{BDF temporal semi-discretization and treatment of non-linearities.}

To produce our BDF-based numerical solver for the
system~\eqref{eq:NavierStokes} we first lay down a semi-discrete
approximation of this equation---discrete in time but continuous in
space---on the basis of the BDF multistep method of order
$s$~\cite[Ch. 3.12]{lambert_numerical_1991}.  Considering the concise
expression~\eqref{eq:GeneralIVP} we let $Q^{j}$ denote the numerical
approximation of $Q$ at time $t=t^j$ and we approximate $Q_t$ at
$t=t^{n+1}$ by the time derivative of the $(d+2)$-dimensional vector
$V=V(t)$ of polynomials of degree $s$ in the variable $t$ that
interpolates the vector values $\{t^{n+1-j},Q^{n+1-j}),\ 0 \leq j \leq
s$. Using the right hand side value $\mathcal{P}(Q^{n+1},t^{n+1})$
this procedure results in the well known implicit $(\Delta t)^{s+1}$
locally-accurate ($(\Delta t)^{s}$ globally-accurate) order-$s$ BDF
formula
\begin{equation} \label{eq:BDF}	
	Q^{n+1} = \sum_{k=0}^{s-1} a_k Q^{n-k} + b \Delta t \, \mathcal{P}(Q^{n+1},t^{n+1}),
\end{equation}
where $a_k$ and $b$ are the BDF coefficients of order
$s$. Table~\ref{table:BDF} displays the BDF coefficients for orders
$s= 1$ through $s=6$.  (The BDF methods of orders greater than six are
not zero-stable~\cite{lambert_numerical_1991}, and therefore do not
converge as $\Delta t\to 0$.)

\begin{table}[!htb]
\centering
\begin{tabular}{c|ccccccc}
	\hline \hline
	$s$ & $a_0$ & $a_1$ & $a_2$ & $a_3$ & $a_4$ & $a_5$ & $b$ \\ \hline
	1 & 1 & & & & & & 1 \\[1.5ex]
	2 & $\frac{4}{3}$ & $-\frac{1}{3}$ & & & & & $\frac{2}{3}$ \\[1.5ex]
	3 & $\frac{18}{11}$ & $-\frac{9}{11}$ & $\frac{2}{11}$ & & & & $\frac{6}{11}$ \\[1.5ex]
	4 & $\frac{48}{25}$ & $-\frac{36}{25}$ & $\frac{16}{25}$ & $-\frac{3}{25}$ & & & $\frac{12}{25}$ \\[1.5ex]
	5 & $\frac{300}{137}$ & $-\frac{300}{137}$ & $\frac{200}{137}$ & $-\frac{75}{137}$ & $\frac{12}{137}$ & & $\frac{60}{137}$ \\[1.5ex]
	6 & $\frac{360}{147}$ & $-\frac{450}{147}$ & $\frac{400}{147}$ & $-\frac{225}{147}$ & $\frac{72}{147}$ & $-\frac{10}{147}$ & $\frac{60}{147}$ \\[1.5ex]
	\hline \hline
\end{tabular}
\caption{Coefficients for BDF methods of orders $s$ with $s=1,\dots,6$.}
\label{table:BDF}
\end{table}

In order to express the resulting algorithm in terms of the
$M$-matrices in equations~\eqref{eq:NavierStokesCurvilinear3D}, for a
given $(d+2)$-dimensional vector valued function $R$ we define the
differential operators
\begin{subequations} \label{eq:Operators} 
\begin{eqnarray}
	\mathcal{A}[R] &=& \sum_{j=0}^2 M^{\xi,j}(R) \frac{\partial^j}{\partial \xi^j} \\
	\mathcal{B}[R] &=& \sum_{j=1}^2 M^{\eta,j}(R) \frac{\partial^j}{\partial \eta^j} \\
	\mathcal{C}[R] &=& \sum_{j=1}^2 M^{\zeta,j}(R) \frac{\partial^j}{\partial \zeta^j} \\
	\mathcal{G}[R] &=& M^{\xi\eta}(R) \frac{\partial^2}{\partial\xi\partial\eta} + M^{\xi\zeta}(R) \frac{\partial^2}{\partial\xi\partial\zeta} + M^{\eta\zeta}(R) \frac{\partial^2}{\partial\eta\partial\zeta},
\end{eqnarray}
\end{subequations} 
in the variables $(\xi,\eta,\zeta)$ (the definition $M^{\xi,0}(R)
\equiv M^0(R)$ was used in these equations, for notational
simplicity). For example, an application of the differential operator
$\mathcal{A}[R]$ to a vector function $S$ results in the expression
\begin{equation}\label{eq:OperatorAFullForm} 
\mathcal{A}[R]S = M^{\xi,0}(R) S + M^{\xi,1}(R) \frac{\partial S}{\partial \xi} + M^{\xi,2}(R) \frac{\partial^2 S}{\partial \xi^2} 
\end{equation}
and similarly for $\mathcal{B}$, $\mathcal{C}$, and $\mathcal{G}$: as
pointed out in Section~\ref{sec:QuasiCartesian}, the $M$ matrices in
equation~\eqref{eq:OperatorAFullForm} generally contain derivatives of the
vector $R$. (As indicated in what follows, the proposed method
produces approximations of the solution $Q$ at time $t=t^{n+1}$ by
taking $R$ and $S$ as suitable---but different---approximations of $Q$
at $t=t^{n+1}$.)  Utilizing the notation~\eqref{eq:Operators},
Equation~\eqref{eq:BDF} can be re-expressed in the form
\begin{equation} \label{eq:UnsplitBDF3D}
	\left( I + b \Delta t\, \mathcal{A} \left[ Q^{n+1} \right] + b \Delta t \, \mathcal{B} \left[ Q^{n+1} \right] + b \Delta t\, \mathcal{C} \left[ Q^{n+1} \right] \right) Q^{n+1} = \sum_{k=0}^{s-1} a_k Q^{n-k} - b \Delta t \, \mathcal{G} \left[ Q^{n+1} \right] Q^{n+1}.
\end{equation}

As mentioned in Section~\ref{sec:Introduction}, previous ADI-based
Navier-Stokes solvers have relied on either linearization or
iterations to adequately account for nonlinear terms. The methods
proposed in this paper, in contrast, utilize the polynomial
extrapolations
\begin{equation} \label{eq:TemporalExtrapolation}
	\widetilde{Q}_p^{n+1} \; \equiv \; \sum_{k = 0}^{p-1} (-1)^k \, {p \choose k + 1} \, Q^{n-k}  \quad (p \geq 1)
\end{equation}
(with $p=s$) to approximate the matrix-valued functions $M$ in the
operators~\eqref{eq:Operators}---which, in particular, gives rise to
{\em high-order-accurate} approximations of the nonlinear terms at
time $t_{n+1}$. (Extrapolation by means of
equation~\eqref{eq:TemporalExtrapolation} with other values of $p$ is
also used as part of the proposed algorithm; see, in particular,
Remark~\ref{rem:OverAccuracy}.) The
formula~\eqref{eq:TemporalExtrapolation} is obtained by evaluating at
$t=t^{n+1}$ the Lagrange interpolating polynomial
$$ \widetilde{Q}_p(t) = \sum_{k = 0}^{p-1} \frac{\ell_k(t)}{\ell_k(t^{n-k})} \, Q^{n-k}, $$
where $t^m = m\,\Delta t$ are equispaced points in time and where
$$ \ell_k(t) = \prod_{\substack{ 0 \leq j \leq p-1 \\ j \neq k }} (t - t^{n-j}). $$
It follows that
$$ \widetilde{Q}_p^{n+1} = Q^{n+1} + \mathcal{O}((\Delta t)^p). $$

Using the extrapolated solution we then proceed as follows: defining
the variable coefficient differential operators
%
%
\begin{equation}\label{eq:ExtrapolatedOperators}
  \mathcal{A}_s = \mathcal{A}[\widetilde{Q}_s^{n+1}], \quad
  \mathcal{B}_s = \mathcal{B}[\widetilde{Q}_s^{n+1}], \quad
  \mathcal{C}_s = \mathcal{C}[\widetilde{Q}_s^{n+1}], \quad
  \mathcal{G}_s = \mathcal{G}[\widetilde{Q}_s^{n+1}],
\end{equation}
we have
\begin{align*}
  \mathcal{A}_s Q^{n+1} &= \mathcal{A}[Q^{n+1}] Q^{n+1} + \mathcal{O}((\Delta t)^s) \\
  \end{align*}
with similar expressions for the other operators
in~\eqref{eq:ExtrapolatedOperators}. We thus obtain the  {\em linear} equation
\begin{equation} \label{eq:UnsplitBDF3DApproximate}
	\left( I + b \Delta t \, \mathcal{A}_s + b \Delta t \, \mathcal{B}_s + b \Delta t \, \mathcal{C}_s \right) Q^{n+1} = \sum_{k=0}^{s-1} a_k Q^{n-k} - b \Delta t \, \mathcal{G}_s Q^{n+1}
\end{equation}
for $Q^{n+1}$. Clearly, these equations are equivalent to the
corresponding $(s+1)$-th order equation~\eqref{eq:UnsplitBDF3D} up to
an error of order $(\Delta t)^{s+1}$, and therefore they themselves
are locally accurate to order $(s+1)$ in time.  Approximations of
order higher than $s$ for the
operators~\eqref{eq:ExtrapolatedOperators} (e.g. approximation of
$\mathcal{A}$, $\mathcal{B}$, $\mathcal{C}$, and $\mathcal{G}$ in
equation~\eqref{eq:UnsplitBDF3D} by $\mathcal{A}_m$, $\mathcal{B}_m$,
$\mathcal{C}_m$, and $\mathcal{G}_m$ respectively with $m > s$) also
preserves the order of the local truncation error, but we have found
the $m=(s+1)$ resulting algorithms to be unstable; cf.
Remark~\ref{rem:OverAccuracy}.

\subsection{ADI factorizations and splittings}

Sections~\ref{sec:DouglasGunn3D}--\ref{sec:BoundaryConditions3D} describe our
application of the Douglas-Gunn method to the semidiscrete linear
high-order scheme~\eqref{eq:UnsplitBDF3DApproximate}. Adequate
treatment of the boundary conditions is a subject of great importance
that is taken up in Section~\ref{sec:BoundaryConditions3D}.

\subsubsection{Application of the Douglas-Gunn method\label{sec:DouglasGunn3D}}

Adding the cross terms $(b\Delta t)^2 (\mathcal{A}_s \mathcal{B}_s +
\mathcal{A}_s \mathcal{C}_s + \mathcal{B}_s \mathcal{C}_s )Q^{n+1}$ and
$(b\Delta t)^3 \mathcal{A}_s \mathcal{B}_s \mathcal{C}_s Q^{n+1}$ to both sides
of equation~\eqref{eq:UnsplitBDF3DApproximate}
and factoring the resulting left-hand side exactly we obtain
\begin{eqnarray} \label{eq:FactoredBDF3D}
	\left( I + b \Delta t \, \mathcal{A}_s \right) \left( I + b \Delta t \, \mathcal{B}_s \right) \left( I + b \Delta t \, \mathcal{C}_s \right) Q^{n+1} &=& \sum_{k=0}^{s-1} a_k Q^{n-k} - b \Delta t \, \mathcal{G}_s Q^{n+1} \nonumber \\
		& & + (b \Delta t)^2 \left( \mathcal{A}_s \mathcal{B}_s + \mathcal{A}_s \mathcal{C}_s + \mathcal{B}_s \mathcal{C}_s \right) Q^{n+1}  \nonumber \\
	    & & + (b \Delta t)^3 \mathcal{A}_s \mathcal{B}_s \mathcal{C}_s Q^{n+1}.
\end{eqnarray}
To eliminate the dependence on $Q^{n+1}$ on the right hand side of
this equation we resort once again to extrapolation: the argument
$Q^{n+1}$ in the right-hand-side term $b\,\Delta t\,\mathcal{G}_s
Q^{n+1}$ is substituted, with error of order $(\Delta t)^{s+1}$, by
the extrapolated value $\widetilde{Q}_s^{n+1}$, and $Q^{n+1}$ in the terms
$(b\Delta t)^2 (\mathcal{A}_s \mathcal{B}_s + \mathcal{A}_s
\mathcal{C}_s + \mathcal{B}_s \mathcal{C}_s )Q^{n+1}$ and $(b\Delta
t)^3 \mathcal{A}_s \mathcal{B}_s \mathcal{C}_s Q^{n+1}$ are
substituted by the extrapolated value $\widetilde{Q}_{s-1}^{n+1}$
(equation~\eqref{eq:TemporalExtrapolation} with $p=s-1$). We thus
obtain the equation
\begin{eqnarray} \label{eq:FactoredBDF3DApprox}
	\left( I + b \Delta t \, \mathcal{A}_s \right) \left( I + b \Delta t \, \mathcal{B}_s \right) \left( I + b \Delta t \, \mathcal{C}_s \right) Q^{n+1} &=& \sum_{k=0}^{s-1} a_k Q^{n-k} - b \Delta t \, \mathcal{M}_s \widetilde{Q}_s^{n+1} \nonumber \\
		& & + (b \Delta t)^2 \left( \mathcal{A}_s \mathcal{B}_s + \mathcal{A}_s \mathcal{C}_s + \mathcal{B}_s \mathcal{C}_s \right) \widetilde{Q}_{s-1}^{n+1}  \nonumber \\
	    & & + (b \Delta t)^3 \mathcal{A}_s \mathcal{B}_s \mathcal{C}_s \, \widetilde{Q}_{s-1}^{n+1}
\end{eqnarray}
whose solution only requires inversion of the {\em linear} operators
$\left( I + b \Delta t \, \mathcal{A}_s \right)$, $\left( I + b \Delta
  t \, \mathcal{B}_s \right)$, and $\left( I + b \Delta t \,
  \mathcal{C}_s \right)$.
\begin{remark}\label{rem:OverAccuracy}
  Notice that, although the approximation $Q_{s-1}^{n+1}$ is accurate
  to order $(s-1)$, the overall accuracy order in quantities such as
  $( b \Delta t )^2 \mathcal{A}_s \mathcal{B}_s
  \widetilde{Q}_{s-1}^{n+1}$, etc., is $(\Delta t)^{s+1}$ as needed---in
  view of the $(\Delta t)^2$ prefactor in this expression. While the
  approximation $Q_{s}^{n+1}$ could have been used while preserving
  the accuracy order, we have found that use of the lower order
  extrapolation $Q_{s-1}^{n+1}$ is necessary to ensure
  stability. Similar comments apply to the term $b \Delta t \,
  \mathcal{G}_s \widetilde{Q}_s^{n+1}$. Note, however, that the term that
  is multiplied by $(\Delta t)^3$ is extrapolated to order $(s-1)$
  rather than $(s-2)$. Although our experiments suggest that using a
  higher order extrapolation than strictly necessary for the terms of
  order $\Delta t$ and $(\Delta t)^2$ could give rise to instability,
  we have found that the extrapolation to order $s-1$ for the term
  multiplied by $(\Delta t)^3$ term does not affect the stability of
  the method, and is therefore used as it gives rise to the relatively
  simple expressions displayed in~\eqref{eq:ADIIn3DRewritten} below.
\end{remark}

To complete the proposed ADI scheme an appropriate splitting of
equation~\eqref{eq:FactoredBDF3DApprox} (that is, an alternating
direction method for evaluation of $Q^{n+1}$) must be used. For
reasons that will become clear in
Sections~\ref{sec:BoundaryConditions3D} and~\ref{sec:BCDiscussion} we use the
Douglas-Gunn
splitting~\cite{douglas_alternating_1962,douglas_two_1963,douglas_jr._general_1964}
and we thus obtain the ADI scheme
\begin{subequations} \label{eq:ADIIn3DRewritten}
\begin{eqnarray}
	\left( I + b \Delta t \, \mathcal{A}_s \right) Q^{*} &=& \sum_{k=0}^{s-1} a_k Q^{n-k} - b \Delta t \, \mathcal{M}_s  \widetilde{Q}_s^{n+1} \nonumber \\
		& & - b \Delta t \, \left( \mathcal{B}_s + \mathcal{C}_s \right) \widetilde{Q}_{s-1}^{n+1} \label{eq:ADIIn3DRewrittenA} \\
	\left( I + b \Delta t \, \mathcal{B}_s \right) Q^{**} &=& \sum_{k=0}^{s-1} a_k Q^{n-k} - b \Delta t \, \mathcal{M}_s \widetilde{Q}_s^{n+1} \nonumber \\
		& & - b \Delta t \, \mathcal{A}_s Q^* -b \Delta t \, \mathcal{C}_s \widetilde{Q}_{s-1}^{n+1} \label{eq:ADIIn3DRewrittenB} \\
	\left( I + b \Delta t \, \mathcal{C}_s \right) Q^{n+1} &=& \sum_{k=0}^{s-1} a_k Q^{n-k} - b \Delta t \, \mathcal{M}_s \widetilde{Q}_s^{n+1} \nonumber \\
		& & - b \Delta t \, \mathcal{A}_s Q^* - b \Delta t \, \mathcal{B}_s Q^{**}. \label{eq:ADIIn3DRewrittenC}
\end{eqnarray}
\end{subequations}
Multiplying~\eqref{eq:ADIIn3DRewrittenC} on the left by $(I + b \Delta
t \mathcal{A}_s)(I + b \Delta t \mathcal{B}_s)$ and eliminating $Q^*$
and $Q^{**}$, it follows that~\eqref{eq:ADIIn3DRewritten} is
equivalent to~\eqref{eq:FactoredBDF3DApprox} up to terms on the order
$\mathcal{O}((\Delta t)^{s+1})$ of the truncation error.

Another form of the Douglas-Gunn splitting is given by
\begin{subequations} \label{eq:ADIIn3D}
\begin{align}
	\left( I + b \Delta t \, \mathcal{A}_s \right) Q^{*} \; =& \; \sum_{k=0}^{s-1} a_k Q^{n-k} - b \Delta t \, \mathcal{M}_s \widetilde{Q}_s^{n+1} \nonumber \\
	   & \; - b \Delta t \, \left( \mathcal{B}_s + \mathcal{C}_s \right) \widetilde{Q}_{s-1}^{n+1} \\
	\left( I + b \Delta t \, \mathcal{B}_s \right) Q^{**} \; =& \; Q^{*} + b \Delta t \, \mathcal{B}_s \widetilde{Q}_{s-1}^{n+1} \\
	\left( I + b \Delta t \, \mathcal{C}_s \right) Q^{n+1} \; =& \; Q^{**} + b \Delta t \, \mathcal{C}_s \widetilde{Q}_{s-1}^{n+1}, \label{eq:ADIIn3DC}
\end{align}
\end{subequations}
which is equivalent to~\eqref{eq:ADIIn3DRewritten} (as it can be
checked by subtracting equations~\eqref{eq:ADIIn3DRewrittenA}
and~\eqref{eq:ADIIn3DRewrittenB}
from~\eqref{eq:ADIIn3DRewrittenC}). The splitting~\eqref{eq:ADIIn3D}
is less expensive than~\eqref{eq:ADIIn3DRewritten}, since
1)~Equation~\eqref{eq:ADIIn3D} does not require computation of the
terms $\mathcal{A}_s Q^*$ and $\mathcal{B}_s Q^{**}$, and 2)~The terms
$b \Delta t \, \mathcal{B}_s \, \widetilde{Q}_{s-1}^{n+1}$ and $b \Delta t
\, \mathcal{C}_s \, \widetilde{Q}_{s-1}^{n+1}$ in~\eqref{eq:ADIIn3D} can
be computed once for each full time-step and used in each ADI sweep as
needed. Therefore the splitting~\eqref{eq:ADIIn3D} is used in the
implementation presented in Section~\ref{sec:NumericalImplementation}.

\subsubsection{Order-preserving boundary conditions for the split
  equations~\eqref{eq:ADIIn3D}} \label{sec:BoundaryConditions3D}

Use of the three-dimensional ADI splitting~\eqref{eq:ADIIn3D} entails
evaluation of solutions of systems of ODEs for the intermediate
unknowns $Q^*$ and $Q^{**}$ as well as the physical unknown $Q^{n+1}$,
each one of which requires enforcement of appropriate boundary
conditions. Here we show that equations~\eqref{eq:ADIIn3DRewritten}
(and thus also~\eqref{eq:ADIIn3D}) possess the following remarkable
property: imposing boundary conditions for $Q^*$ and $Q^{**}$ which
coincide with the corresponding boundary conditions for $Q$ at time
$t=t^{n+1}$ preserves the overall $(\Delta t)^{s+1}$ truncation error
otherwise implicit in these equations.

In view of Remark~\ref{rem:InflowBC}, in what follows we assume the
Navier-Stokes boundary conditions 
\begin{equation} \label{eq:BoundaryConditions3D}
\begin{pmatrix}
	\mathbf{u}(\xi,\eta,\zeta,t) \\ T(\xi,\eta,\zeta,t)
\end{pmatrix}
=
\begin{pmatrix}
	g^\mathbf{u}(\xi,\eta,\zeta,t) \\ g^T(\xi,\eta,\zeta,t)
\end{pmatrix}
,
\quad (\xi,\eta,\zeta) \in \partial D,
\end{equation}
for the unknown $Q = (\mathbf{u}^{\mathrm T},T,\rho)^{\mathrm T} \,
\in \, \mathbb{R}^{3+2}$ ($Q = Q(\xi,\eta,\zeta,t)$) at a solid-fluid
interface. Comparison of equations~\eqref{eq:ADIIn3DRewrittenC}
and~\eqref{eq:UnsplitBDF3DApproximate} shows that the truncation error
in~\eqref{eq:ADIIn3DRewrittenC} is a quantity of order $\Delta
t^{s+1}$ provided $Q^*$ and $Q^{**}$ are $s$-order-accurate
approximations of $Q^{n+1}$ everywhere in the domain $D$ and its
boundary. But, as discussed in Section~\ref{sec:BCDiscussion}, the
intermediate unknowns $Q^*$ and $Q^{**}$ are indeed accurate to order
$s$ throughout $D$ provided the boundary conditions of $Q^*$ and
$Q^{**}$ are taken to coincide with those for $ Q(t^{n+1})$.  Thus,
use of boundary values of the solution at time $t=t^{n+1}$ for the
intermediate-time unknowns $Q^*= (\mathbf{u^*}^{\mathrm
  T},T^*,\rho^*)^{\mathrm T}$ and $Q^{**}= (\mathbf{u^{**}}^{\mathrm
  T},T^{**},\rho^{**})^{\mathrm T}$, that is
\begin{subequations}\label{eq:ADIBC3D}
\begin{align}
	\begin{pmatrix}
		\mathbf{u}^*(\xi,\eta,\zeta) \\ T^*(\xi,\eta,\zeta)
	\end{pmatrix}
	=
	\begin{pmatrix}
		g^\mathbf{u}(\xi,\eta,\zeta,t^{n+1}) \\ g^T(\xi,\eta,\zeta,t^{n+1})
              \end{pmatrix},
              & \quad\mbox{for $\xi=\ell_1,\ell_2$ and $\eta,\zeta \in [\ell_1,\ell_2]$} \label{eq:QstarBC3D} \\
	\begin{pmatrix}
		\mathbf{u}^{**}(\xi,\eta,\zeta) \\ T^{**}(\xi,\eta,\zeta)
	\end{pmatrix}
	=
	\begin{pmatrix}
		g^\mathbf{u}(\xi,\eta,\zeta,t^{n+1}) \\ g^T(\xi,\eta,\zeta,t^{n+1})
              \end{pmatrix},
              & \quad\mbox{for $\eta=\ell_1,\ell_2$ and $\xi,\zeta \in [\ell_1,\ell_2]$} \label{eq:QstarstarBC3D} \\
	\begin{pmatrix}
		\mathbf{u}^{n+1}(\xi,\eta,\zeta) \\ T^{n+1}(\xi,\eta,\zeta)
	\end{pmatrix}
	=
	\begin{pmatrix}
		g^\mathbf{u}(\xi,\eta,\zeta,t^{n+1}) \\ g^T(\xi,\eta,\zeta,t^{n+1})
	\end{pmatrix},
	& \quad\mbox{for $\zeta=\ell_1,\ell_2$ and $\xi,\eta \in [\ell_1,\ell_2]$},
\end{align}
\end{subequations}
maintains the overall $\mathcal{O}(\Delta t)^{s+1}$ truncation error
in the Douglas-Gunn scheme for the complete time-step $t^n\to
t^{n+1}$. The results in Section
\ref{sec:NumericalResults} demonstrate the expected order of accuracy
is achieved in the case of general boundary conditions, including
cases in which time-dependent boundary conditions are specified.

\subsection{Discussion: enforcement of boundary conditions in the
  present and previous ADI schemes} \label{sec:BCDiscussion}

This section provides a justification for our use of the boundary
values of $Q(t^{n+1})$ in the solution of the intermediate equations,
and it highlights the advantages of the present strategy over other
methods for enforcement of boundary conditions in ADI schemes.

As discussed in~\cite{leveque_intermediate_1985}, use of given
boundary values as boundary conditions for the intermediate
(unphysical) variables may lead to reductions in the order of accuracy
of the overall solver unless the intermediate time-steps in the ADI
scheme satisfy certain accuracy conditions. The desired full-step
order $\mathcal{O}((\Delta t)^{s+1})$ of temporal accuracy can be
guaranteed provided $Q^*$ and $Q^{**}$ and associated boundary
conditions satisfy certain ``modified'' PDEs of the form
\begin{equation} \label{eq:ModifiedPDE}
  Q^*_t = L_1^* Q^*\quad\mbox{and}\quad Q^{**}_t = L_2^* Q^{**}
\end{equation}
with truncation errors of order $\mathcal{O}((\Delta t)^{s})$
throughout the domain and up to and including the boundary (in view of
the $\Delta t$ factors in equation~\eqref{eq:ADIIn3DRewrittenC}; see
also~\cite[pp. 3--4]{leveque_intermediate_1985}).  Briefly, for
example, the modified differential equation associated with $Q^*$ is
that which {\em would} be solved up to a truncation error of order
$s+1$ via sole iteration of the first intermediate time-stepping
scheme~\eqref{eq:ADIIn3DRewrittenA}.

As indicated in~\cite{leveque_intermediate_1985} the necessary
adequately-accurate boundary conditions for $Q^*$, for example, can be
obtained by applying the Taylor series procedure to the first equation
in~\eqref{eq:ModifiedPDE} with initial conditions $Q^* = Q^n$ at
$t=t^{n}$, to obtain a solution in the form of a truncated series in
powers of $\Delta t$ of the appropriate order which can then be
evaluated at $t=t^{n+1}$ to produce the desired $\mathcal{O}((\Delta
t)^{s})$-accurate boundary condition for $Q^*$ at that time.  This
prescription ensures~\cite{leveque_intermediate_1985} that the errors
in boundary values for intermediate variables $Q^*$ are quantities of
the appropriate order of time-accuracy. It follows that, when a
similar procedure is completed with $Q^{**}$ and when ultimately
$Q^{n+1}$ is computed, agreement between the full-step split and
unsplit discrete schemes to order $(\Delta t)^{s}$ takes place
throughout the domain, up to and including the domain boundary.

The order-$s$ boundary-condition prescriptions provided
in~\cite{leveque_intermediate_1985} are expressed in terms of certain
spatial derivatives of the numerical solution $Q^n$: each subsequent
order of time-accuracy requires an additional term in the formal power
series solution, and, thus, in view of the Taylor series method used,
it requires several spatial derivatives of the numerical solution
$Q^n$ at the boundary. For the heat equation and the Navier-Stokes
equation, for example, two additional spatial derivatives of the
numerical solution at the boundary are in principle necessary for each
additional order of time-accuracy. But, as pointed out
in~\cite{leveque_intermediate_1985}, the original PDE can be used to
express such derivatives in terms of derivatives with respect to time
together with a derivative of the highest order along the boundary
(which do not present difficulties as they can be obtained from the
boundary conditions) as well as numerical derivatives of the discrete
solution $Q^n$ of order lower than the maximum order of spatial
differentiation in the original PDE. In some cases, simplifications
can be made such that the resulting expression for $Q^*$ at the
boundary is given in terms of the given boundary data for $Q$ at
$t=t^n$ and $t^{n+1}$ only.  For the Navier-Stokes equations, however,
evaluation of the boundary condition for $Q^*$ requires
differentiation of the numerical solution at the boundary of orders as
high as the desired order $s$ of temporal accuracy---which could give
rise to accuracy losses and, owing to its dependence on solution
values at time $t=t^n$, may give rise to CFL-type constraints in
otherwise unconditionally stable implicit solvers.

The Douglas-Gunn scheme under consideration is exceptional in that a
modified PDE can be obtained simply and without any recourse to
differentiation: the $t=t^{n+1}$ boundary condition implied by this
modified PDE for the intermediate variables exactly coincides with the
physical boundary conditions at time $t=t^{n+1}$---for any
time-accuracy order used.  In detail, in the Douglas-Gunn scheme the
intermediate relation~\eqref{eq:ADIIn3DRewrittenA} is an approximation
of the full unfactored scheme~\eqref{eq:UnsplitBDF3DApproximate} with
truncation error of order $(\Delta t)^{s}$. (Note that in our scheme
this error arises from the $(s-1)$-order extrapolation used which, as
discussed in Remark~\ref{rem:OverAccuracy}, must be used.) Since $Q^*$
and $Q^{**}$ are multiplied by $ \Delta t$ on the right-hand side of
equation~\eqref{eq:ADIIn3DRewrittenC} this $\mathcal{O}((\Delta
t)^{s})$ additional error does not change the order $\Delta t^{s+1}$
of the truncation error of the scheme for the overall time-step from
$t^n$ to $t^{n+1}$. The aforementioned Taylor expansion procedure
applied at a boundary point and at time $t^{n}$ provides solutions at
time $t^{n+1}$ which, in view of the (assumed) smoothness of solutions
and prescribed boundary data for the original
equation~\eqref{eq:GeneralIVP}, must satisfy the boundary conditions
imposed on the exact solution $Q$ up to an error of the relevant order
$(\Delta t)^s)$. Thus, as claimed, use of $t=t^{n+1}$
physical boundary conditions for evaluation of the intermediate
unknowns $Q^*$ and $Q^{**}$ within the Douglas-Gunn scheme preserves
the overall $\mathcal{O}((\Delta t)^{s+1})$ truncation error of the
original unsplit scheme.
\begin{remark}
  It is interesting to note that {\em for any type of boundary
    conditions imposed on} $Q^{n+1}$, whether of Dirichlet type,
  Neumann type, Robin type, etc., the boundary conditions for the
  intermediate variables $Q^*$ and $Q^{**}$ in the Douglas-Gunn scheme
  necessarily coincide with those imposed on the exact solution at
  time $t^{n+1}$ up to an error of order $(\Delta t)^s$---as required
  to maintain the overall $(\Delta t)^{s+1}$ truncation error. Indeed,
  the exact physical solution $Q$ is a solution of the modified PDE
  for $Q^*$ with an error of the order $(\Delta t)^s$ for $t^n\leq t
  \leq t^{n+1} = t^n + \Delta t$ and, thus, in the present
  semi-discrete context, the full Taylor series in space and time must
  coincide, up to order $(\Delta t)^s$ and for all orders in the
  spatial variables not only at the boundary but {\em throughout the
    physical domain}. Therefore, discrete solutions arising from a
  subsequent spatial discretization satisfy the prescribed boundary
  condition up to error of order $(\Delta t)^s$ and up to the selected
  spatial discretization error at each boundary discretization point.
\end{remark}

\section{BDF-ADI unconditional stability ($s=2$) and quasi-unconditional
  stability ($3\leq s \leq 6$)\label{sec:StabilityDiscussion}}

In lieu of a full stability analysis for the non-linear compressible
Navier-Stokes equations under consideration (for which rigorous
mathematical discussions of stability are not available for any the
various extant algorithms), rigorous stability results for linear
related problems and numerical experiments demonstrating stability
properties for the fully nonlinear cases are presented in Part~II. The
next two sections briefly summarize the results put forth in that
reference. In particular, Section~\ref{sec:StabilityBDF2-ADI} discusses the
stability properties (unconditional stability) of the second order BDF
schemes introduced in Section~\ref{sec:BDF-ADI} above specialized to
the convection and parabolic equations under various
discretizations. Section~\ref{sec:BDFQuasiUnconditional} then introduces the concept
of quasi-unconditional stability and it reviews relevant stability
proofs and results of numerical experiments presented in
Part~II---including rigorous stability proofs for non-ADI BDF-based
algorithms for convection-diffusion equations and numerical evidence
supporting the suggestion that the BDF-ADI based schemes for the
Navier-Stokes equations do in fact exhibit quasi-unconditional
stability.

\subsection{Unconditional stability of the second-order BDF-ADI
  algorithms for the convection and parabolic
  equations\label{sec:StabilityBDF2-ADI}}

As indicated above, Part~II includes proofs of unconditional stability
for the BDF-ADI scheme of order two for linear constant-coefficient
hyperbolic and parabolic equations in two spatial dimensions under
periodic boundary conditions and Fourier spatial discretization.  A
corresponding proof is also presented in that reference for the
non-periodic parabolic equation on the basis of Legendre spatial
discretization. The following theorem summarizes these results in some
detail.

\begin{theorem}\label{thm:BDF2-ADI}
  The second-order BDF-ADI method with periodic boundary conditions
  and Fourier spectral discretization is unconditionally stable for
  both the two-dimensional advection equation
$$
  U_t + \alpha U_x + \beta U_y = 0
$$
(with real constants $\alpha$ and $\beta$) and the parabolic equation
\begin{equation}\label{eq:ParabolicPDE}
  U_t = a U_{xx} + b U_{yy} + c U_{xy}
\end{equation}
(with constants $a,b>0$ and $c$ satisfying $c^2\leq4ab$). Similarly,
the BDF2-ADI method for equation~\eqref{eq:ParabolicPDE} with homogeneous
boundary conditions and using Legendre spectral discretization is
unconditionally stable. In particular, in all of these cases the
energy of the approximate solution $u^n$ at any time step $n\geq 2$
with time-step size $\Delta t$ can be estimated in terms of the first
two time-steps in the solution: we have
$$
  |u^n|^2 \leq C( |u^0|^2 + |u^1|^2 )
$$
for some constant $C$, where $|\cdot|$ denotes the discrete $L^2$
norm.
\end{theorem}

\subsection{Quasi-unconditional stability of non-ADI order-$s$
  BDF-based algorithms for the convection-diffusion
  equation\label{sec:BDFQuasiUnconditional}}

\begin{figure}[!htb]
	\centering
	\includegraphics[width=0.5\textwidth]{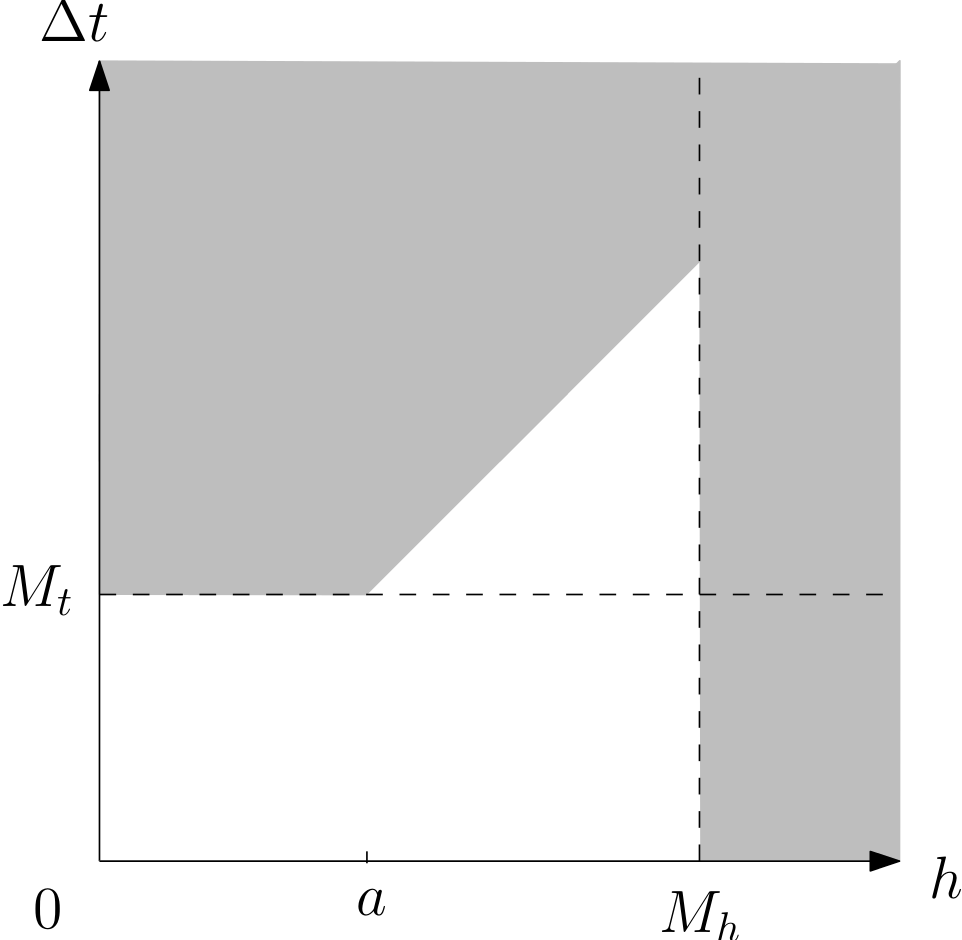}
        \caption{Stability region of a notional quasi-unconditionally
          stable PDE solver is shown in white; the grey region is the
          set of $h$ and $\Delta t$ where the method is unstable. In
          particular, the method is stable in the rectangular region
          $0<h<M_h$, $0<\Delta t<M_t$. Notice that outside of this
          region the method is stable for time steps satisfying
          certain CFL-like contraints (linear in this
          illustration). Quasi-unconditional stability does not
          exclude the possibility of other stability constraints
          outside of the rectangular region of stability.}
	\label{fig:QuasiUnconStab}
\end{figure}

As is well known, the BDF schemes of order $s\geq 3$ are not A-stable
(as it follows, for example, from the well known Dahlquist's second
barrier~\cite{dahlquist_special_1963}, which states that any
multi-step A-stable method must be at most second order
accurate). However, a certain concept of \emph{quasi-unconditional
  stability} emerges in the context of the proposed high-order BDF-ADI
solvers. In detail, denoting by $\Delta t$ the temporal time-step and
letting $h$ denote a parameter that controls the spatial meshsize, we
introduce the following definition.
\begin{definition}\label{def:QuasiUnconditional}
  A numerical method for the solution of the PDE $Q_t =
  \mathcal{P}\,Q$ in $\Omega$ is said to be
  \textbf{\emph{quasi-unconditionally stable}} if there exist positive
  constants $M_h$ and $M_t$ (which generally depend on the physical
  parameters, initial conditions and boundary conditions) such that
  the method is stable for all $h < M_h$ and all $\Delta t < M_t$.
\end{definition}
In other words, a quasi-unconditionally stable solver possesses the
following property: for each domain $\Omega$ and each selection of
boundary and initial conditions and source terms there exists a {\em
  fixed} threshold $M_t$ such that, for each $\Delta t< M_t$ the
solver is stable for arbitrarily small spatial mesh-sizes.  Note that
other stability constraints might hold outside of the
quasi-unconditional stability rectangle $(0,M_h)\times(0,M_t)$.
Figure~\ref{fig:QuasiUnconStab} illustrates the concept of
quasi-unconditional stability in the parameter space $(h,\Delta t)$ in
a case where a CFL type constraint exists outside the rectangle
$(0,M_h)\times(0,M_t)$.  Part~II establishes rigorously the
quasi-unconditional stability of BDF methods of orders $3\leq s\leq 6$
for convection-diffusion equations, and it presents results of
numerical tests which clearly suggest the proposed BDF-ADI methods for
the Navier-Stokes equations enjoy this property as well. The
aforementioned rigorous results are summarized in the following
theorem.
\begin{theorem}\label{thm:BDFAdvDiff}
  The BDF methods of order $s$, $2 \leq s \leq 6$ (no ADI!) with
  periodic boundary conditions and Fourier spectral discretization are
  quasi-unconditionally stable for the constant-coefficient
  advection-diffusion equation
$$
U_t + \boldsymbol \alpha \cdot \nabla U = \beta \Delta U
$$
($\alpha \in \mathbb R^d$, $\beta>0$) in $d=1$, $2$, and $3$
dimensional space. The corresponding constants $M_h$ and $M_t$ in
Definition~\ref{def:QuasiUnconditional} are given by $M_h=\infty$ and
$M_t = \frac{\beta}{|\boldsymbol \alpha|^2} m_C$, where $m_C$ is an
explicitly computable constant depending on the order $s$ of the
method which is independent of physical parameters, initial conditions
and boundary conditions. The values of the constant $m_C$ are listed
in Table~\ref{table:ConstantM_C}.

\begin{table}[!htb]
\centering
\begin{tabular}{c|cccc}
\hline \hline
 $s$ & 3 & 4 & 5 & 6 \\ \hline
 $m_C$ & 14.0 & 5.12 & 1.93 & 0.191 \\ \hline \hline
\end{tabular}
\caption{Numerical values of the constant $m_C$ such that the order-$s$  BDF method applied to the advection-diffusion equation $u_t + \alpha \, u_x = \beta \,u_{xx}$ with Fourier collocation is stable for all $\Delta t < \frac{\beta}{\alpha^2}m_C$ and for all $h>0$.}
\label{table:ConstantM_C}
\end{table}

\end{theorem}

As mentioned above, a quasi-unconditionally stable algorithm can be
stable even in cases in which the constraints on $(h,\Delta t)$ in
Definition~\ref{def:QuasiUnconditional} are not satisfied, and, in
such cases, stability may still take place under certain CFL-like
conditions; see Part~II for details. Briefly, that reference presents
tabular data which display numerically estimated maximum stable values
of $\Delta t$ for two- and three-dimensional Navier-Stokes BDF-ADI
methods and for a number of spatial-discretization sizes. This data
suggests clearly that, for each $s$, maximum stable $\Delta t$ values
do approach positive asymptotic limits as finer and finer spatial
discretizations are used, as befits a quasi-unconditionally stable
scheme.

\section{Numerical Implementation\label{sec:NumericalImplementation}}

In this section we present details of our full spatio-temporal
implementation of the BDF-ADI algorithms discussed above in this text.
Spatial discretizations of various kinds can be used in these
contexts, including finite-difference, polynomial-spectral and
Fourier-continuation~\cite{bruno_higher-order_2014}
discretizations. For the sake of definiteness we restrict our
presentation to the Chebyshev-collocation spatial
approximation~\cite{kopriva_implementing_2009,boyd_chebyshev_2001}
which is briefly reviewed in the following section; results arising
from use of the Fourier spectral
method~\cite{kopriva_implementing_2009,boyd_chebyshev_2001} are also
included in Section~\ref{sec:NumericalResults}.

\subsection{Chebyshev spatial discretization, GMRES iterations,
  spectral filtering, geometrical metric terms\label{sec:DiscreteSpatialDetails}}

Let the computational PDE domain $D=[-1,1]^3$ be discretized by means
of an $(N+1)$-node Gauss-Lobatto Chebyshev
discretization~\cite{kopriva_implementing_2009,boyd_chebyshev_2001} in
the $\xi$, $\eta$, and $\zeta$ directions:
\begin{eqnarray}
	\xi_i = -\cos(\pi i / N_\xi) &,& i=0,\dots,N_\xi, \nonumber \\
	\eta_j = -\cos(\pi j / N) &,& j=0,\dots,N_\eta,\nonumber \\
	\zeta_k = -\cos(\pi k / N) &,& k=0,\dots,N_\zeta;\nonumber
\end{eqnarray}
the treatment for the two-dimensional case is, of course, entirely
analogous.  In what follows the PDE solution $Q$ is approximated
numerically by means of grid discretizations $Q_{ijk} \sim
Q(\xi_i,\eta_j,\zeta_k)$ together with the associated Chebyshev
expansions
\begin{equation} \label{eq:FourierInterpolant}
  Q_N(\xi,\eta,\zeta) = \sum_{{\bf 0}\leq (i,j,k)\leq N} \widehat{Q}_{ijk} T_i(\xi) T_j(\eta) T_k(\zeta),
\end{equation}
in terms of the Chebyshev polynomials $T_\ell$ ($\ell\geq 0$), where
$N=(N_\xi,N_\eta,N_\zeta)$, where $\bf 0$ denotes the
three-dimensional zero vector, and where inequalities between
three-dimensional vectors are interpreted in component-wise
fashion. The coefficients $\widehat Q_{ijk}$ of the numerical
approximation $Q_N$ are related to the point values $Q_{ijk}$ by the
interpolation relations $Q_N(\xi_i,\eta_j,\zeta_k) = Q_{ijk}$ (${\bf
  0}\leq (i,j,k)\leq N$).

The discrete Chebyshev spatial differentiation operators we use are
standard~\cite{kopriva_implementing_2009,boyd_chebyshev_2001}: the
$\xi$-derivative operator $\delta_\xi$ applied to a grid function
$Q_{ijk}$, for example, is defined as the grid function
$(\delta_\xi Q)_{ijk}$ whose $(ijk)$ value equals the value of the
derivative of the interpolant $Q_N$ at the point
$(\xi_i,\eta_j,\zeta_k)$:
\begin{equation} \label{eq:DerivativeExact} 
  (\delta_\xi Q)_{ijk} = \frac{\partial}{\partial \xi} Q_N(\xi_i,\eta_j,\zeta_k). 
\end{equation}
Similar definitions are used for the operators $\delta_{\xi\xi}$,
$\delta_\eta$, $\delta_{\eta\eta}$, $\delta_{\xi\eta}= \delta_\xi
\delta_\eta = \delta_\eta \delta_\xi$ etc. As is common practice, for
all of the numerical examples presented in this paper the Chebyshev
derivatives are evaluated efficiently by means of the fast cosine
transform.



Using the Chebyshev discretizations mentioned above, the
one-dimensional boundary value problems given by the ODE
systems~\eqref{eq:ADIIn3D} and the boundary
conditions~\eqref{eq:ADIBC3D} become discrete systems of linear
equations. In order to fully take advantage of the fast cosine
transform we solve these systems by means of the GMRES iterative
solver with second order finite difference preconditioner
(cf.~\cite[p. 293]{boyd_chebyshev_2001}
and~\cite{bruno_spatially_2014}); a similar treatment is used in
conjunction with Fourier spectral discretizations.

For both Chebyshev and Fourier-spectral discretizations an exponential
filter~\cite{gottlieb_gibbs_1997}, which does not degrade the $s$-th
order accuracy of the method
(cf.~\cite[Sec. 4.3]{albin_spectral_2011}), is employed to ensure
stability.  In the Chebyshev case, for example, the filtered
coefficients $h^f_k$ for a given function $h=\sum_k \widehat h_k T_k(x)$
are given by
$$ h^f_k = \exp \left( -\alpha \left( \frac{k}{N} \right)^{2p} \right)
\widehat h_k.$$
For all of the results presented in this paper we have set $\alpha =
16 \log 10$ and $p = 8$. The filter is applied at the end of the time
step to each line of discretization points in each dimension,
requiring one fast cosine transform per line to obtain the
coefficients $\widehat h_k$, and one transform to obtain the filtered
physical function values.

The transformation of the equations to general coordinates requires
the metric terms $\xi_x$, $\xi_y$, etc; see
Section~\ref{sec:QuasiCurvilinear}. The solvers presented in this
paper use the so-called ``invariant form'' of these metric
terms~\cite{thomas_navier-stokes_1990}, but other (accurate)
alternatives could be equally advantageous. The derivatives of the
physical coordinates ($x_\xi$, $x_\eta$, etc.) needed in the actual
expressions for the metric terms are produced by means of the discrete
derivative operators implicit in the Chebyshev or Fourier spatial
approximation used in each case.

\subsection{Overall algorithmic description and treatment of boundary
  values and initial time-steps\label{sec:BCImplementation}}

Given the elements described in previous sections of this paper, our
actual implementations of BDF-ADI algorithms of order $s$ can now be
described in rather simple terms---except perhaps for some details,
which require additional considerations, concerning boundary values of
the fluid density and the possible presence of corners and edges in
the boundary of the computational domain.  The absence of a density
boundary condition has previously been successfully addressed by means
of discretization strategies based on use of staggered grids see
e.g.~\cite[Ch. 4.6]{canuto_spectral_2007} and the references
therein. In some such strategies the velocity and the temperature are
collocated on a Gauss-Lobatto grid while the density is collocated on
a Gauss grid---so that the density mesh contains no boundary points,
and therefore no density boundary conditions are needed. In the
context of ADI-based methods such as the ones considered in this
paper, however, it is not clear that a natural staggered-grid ADI
method could be designed---since the ADI approach requires solution of
one-dimensional boundary value problems which couple all field
components. An alternative approach is proposed in this paper. This
method uses the same Gauss-Lobatto grid for all unknowns, including
the density, and therefore it requires determination of the boundary
values of the density as part of the overall solution.

Our approach in these regards follows from the following observation:
the density components of the unknowns $Q^*$, $Q^{**}$ and $Q^{n+1}$
throughout the domain and including the boundary can be obtained by
interpreting the corresponding equations~\eqref{eq:ADIIn3DRewritten}
(or, equivalently, equations~\eqref{eq:ADIIn3D}) as a system which
includes the density boundary values as unknowns. We demonstrate the
method in detail in the case of equation~\eqref{eq:ADIIn3DRewrittenA};
the treatment of the other equations in
either~\eqref{eq:ADIIn3DRewritten} or~\eqref{eq:ADIIn3D} is
analogous. As suggested above, for a fixed pair $(k,\ell)$ we view
equation~\eqref{eq:ADIIn3DRewrittenA} as a relation between
$(d+2)(N-1)+2$ unknowns ($d=3$ in the present example), namely, the
discrete values $Q^*_{jk\ell}$ of the vector $Q^*=
((\mathbf{u}^*)^{\mathrm T},T^*,\rho^*)^{\mathrm T}$ that corresponds
to discretization points in the interior of the PDE domain ($1\leq j
\leq N-1$) {\em together with} the density boundary values
($\rho^{n+1}_{jk\ell}$ for $j=0$ and $j=N$). Clearly, collocation
of~\eqref{eq:ADIIn3DRewrittenA} at all interior points along the
$(k,\ell)$ discretization segment furnishes $(d+2)(N-1)$ equations for
these unknowns. 

The necessary two additional equations are obtained by enforcing the
portion of~\eqref{eq:ADIIn3DRewrittenA} that arises from the mass
conservation equation at each one of the two boundary points $j=0$ and
$j=N$. Note that in order to solve the overall system of
$(d+2)(N-1)+2$ equations along the $(k,\ell)$ discretization segment
at time $t^{n+1}$, the values of $Q= (\mathbf{u}^{\mathrm
  T},T,\rho)^{\mathrm T}$ at the corresponding interior discretization
points and boundary points together with the boundary values of
$\mathbf u$ and $T$ (given by the boundary
conditions~\eqref{eq:QstarBC3D}) at the boundary points must be
available for all time-steps $t^m$ with $n-s+1\leq m\leq n$. (For the
subsequent equations in~\eqref{eq:ADIIn3DRewritten}
or~\eqref{eq:ADIIn3D} the values of $Q^*$ and $Q^{**}$ evaluated in
the previous intermediate steps at interior and boundary points are
needed as well.) Using such data the algorithm produces the needed
interior and boundary values $Q^*_{jk\ell}$. As mentioned above, the
subsequent equations for the unknowns $Q^{**}$ and $Q^{n+1}$
in~\eqref{eq:ADIIn3DRewritten} or~\eqref{eq:ADIIn3D} are treated
similarly.

It is important to note that in the algorithm just described, the
third ADI sweep produces not only the values of the density
$\rho^{n+1}$ at the horizontal boundary faces $\zeta =\ell_1$ and
$\zeta =\ell_2$, but also the values of $\rho^{n+1}$ along the
vertical boundary faces $\xi=\ell_1$, $\xi=\ell_2$, $\eta=\ell_1$ and
$\eta=\ell_2$. These are necessary to form the BDF-ADI
system~\eqref{eq:ADIIn3DRewritten} at subsequent time-steps. However,
as part of the solution along the vertical faces, values for $\mathbf
u^{n+1}$ and $T^{n+1}$ are also produced, and we have observed that
retaining these values for the solution causes a degradation in the
order of accuracy (although the solvers continue to enjoy
quasi-unconditional stability in this case).  Therefore, in order to
preserve the order $s$ of time-accuracy, it is important to discard
these values of $\mathbf u$ and $T$ and substitute them by those given
by the boundary conditions~\eqref{eq:ADIBC3D}. This completes the
description of the algorithm.

We emphasize that no special boundary conditions are required for
either the intermediate density $\rho^*$ or the final density
$\rho^{n+1}$. The density is determined entirely by
equations~\eqref{eq:ADIIn3D} throughout the domain, up to and
including the boundary.  Furthermore, the presence of corners does not
impact the stability of the solver: no special boundary treatment for
the corners of the domain are necessary.

\begin{remark}\label{rem:Corners}
  Although corners in the computational domain do not affect the
  stability of our solvers, we note that the spatial accuracy may
  deteriorate as a result of singularities that occur at corners and
  edges of the computational domain (see
  e.g.~\cite[Ch. 6.12]{boyd_chebyshev_2001} for a corresponding
  discussion in the context of for spectral discretizations). Provided
  the physical domain contains no corner or edges, this problem can be
  eliminated by means of a multi-domain overset
  decomposition~\cite{brown_overture:_1999,cubillos_general-domain_2015},
  in such a way that actual PDE solutions around corner regions and
  edge regions in one computational patch are actually replaced by
  solution values obtained at interior regions in other
  patches. Actual physical corners and edges, which also give rise to
  accuracy reductions, can be treated in a variety of ways, but such
  considerations are beyond the scope of this paper. In any case, the
  numerical examples in the next section show the correct order of
  time-accuracy of the solvers with a manufactured solution in a
  computational domain containing corners and edges, as well as a
  physical solution in an annular domain---which, of course, contains
  no corners.
\end{remark}

To conclude this section we provide some comments concerning
evaluation of solution values at the first $s$ timesteps in a method
of overall accuracy order $s$.  In the simplest approach the solution
is ramped-up from a constant field state (usually zero for all
velocities and one for the density and temperature), and the
simulation is arranged in such a way that the solution values at each
one of the initial $s$ time-steps is known and equal to the assumed
constant state. But in some situations evaluation of the transients
from given initial conditions may need to be obtained; see e.g. the
example provided in Section~\ref{sec:NumericalResults} involving flow
in an annulus, where the density has a non-constant initial
condition. Use of explicit solvers is some times recommended to obtain
the first few solution values, but such explicit solvers generally
require use of significantly smaller time-steps than those used by the
implicit solver---in view of their inherent properties of conditional
stability.  Furthermore, a high order multi-step explicit solver would
also require previous time levels, and a Runge-Kutta method requires
special treatment of boundary conditions for the intermediate
stages. In order to avoid such difficulties, we propose a strategy
based on use of the first-order BDF-ADI method followed by
Richardson-extrapolation (cf.~\cite{lyon_high-order_2010} and
references therein) of a sufficiently high order so as to match the
overall order of time-accuracy of the method. For example, to produce
a second-order accurate solution at $t=\Delta t$ from initial data
$Q^0$, two solutions using the first-order BDF-ADI algorithm are
computed at time $t=\Delta t$---one solution ($Q^1_{(1)}$) with time
step equal to $\Delta t$, the other ($Q^1_{(2)}$) with time-step
$\Delta t/2$. The second-order accurate solution $Q^1$ is obtained as
an appropriate linear combination of $Q^1_{(1)}$ and $Q^2_{(2)}$: $
Q^1 = 2Q^1_{(2)}-Q^1_{(1)}. $

\section{Numerical Results \label{sec:NumericalResults}}

This section presents a variety of numerical results produced by the
BDF-ADI solvers introduced in this paper for two- and
three-dimensional spatial domains and for orders $s$ with $2\leq s\leq
6$. In particular, these results demonstrate that the proposed solvers
do enjoy the claimed spatial and temporal orders of accuracy and
general applicability; detailed studies demonstrating the claimed
stability properties are deferred to Part~II. All of the numerical
examples were obtained from runs on either a single core of an Intel
i5-2520M processor with 4 GB of memory, or a single core of an Intel
Xeon X5650 processor with 24 GB of memory.  Unless otherwise
indicated, all simulations use the parameter values $\mathrm{Pr}=0.71$
and $\gamma=1.4$, and the (non-dimensional) viscosity and thermal
conductivity are given by Sutherland's law~\eqref{eq:Sutherland} with
$S_\kappa=S_\mu=0.3$.

\begin{figure}[!htb]
	\centering
	\includegraphics[width=0.5\textwidth]{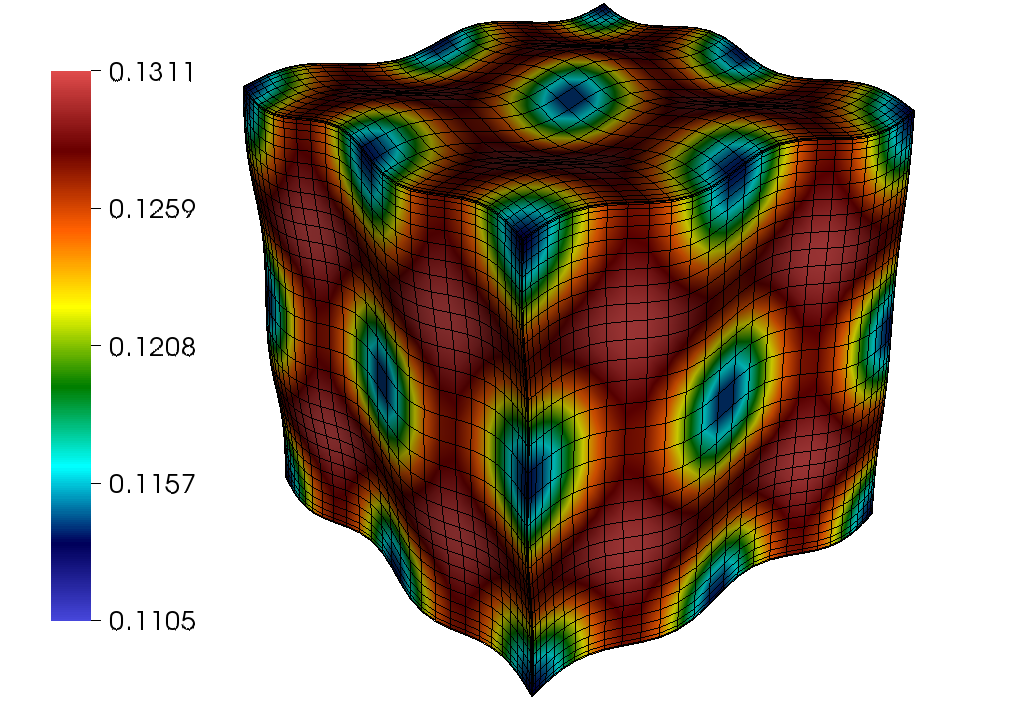}
	\caption{The wavy cube used in the convergence tests of the
          three-dimensional solver. The coloring corresponds to the
          Jacobian of transformation.}
	\label{fig:WavyCube}
\end{figure}

\begin{figure}[!htb]
	\centering
	\includegraphics[width=0.75\textwidth]{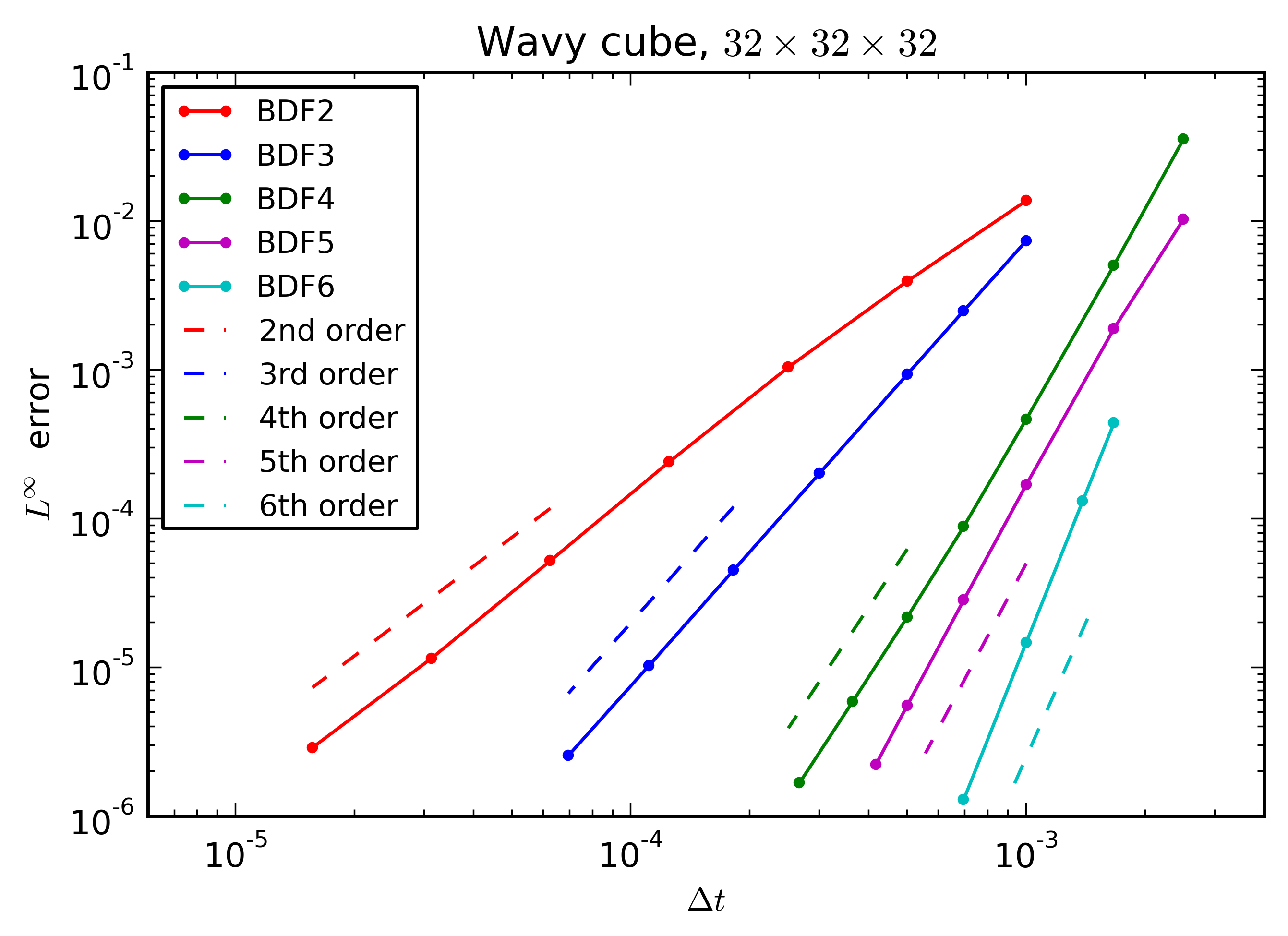}
	\caption{Convergence of the three-dimensional BDF-ADI solvers
          of orders $s$, $s=2,\dots,6$, for the Navier-Stokes problem
          in the wavy cube.}
	\label{fig:ManufConv3DWavy}
\end{figure}

\begin{table}[!htb]
\centering
\begin{tabular}{c|ccccccc}
	\hline \hline
	$Q_j$ & $\alpha_j$ & $\beta_j$ & $\lambda_j$ & $\phi_{j,t}$ & $\phi_{j,\xi}$ & $\phi_{j,\eta}$ & $\phi_{j,\zeta}$ \\
	\hline
	$u$    & 0 & 1   & 25 & -1 & 0 & 0 & 0 \\ 
	$v$    & 0 & 1   & 25 & -2 & 0 & 0 & 0 \\ 
	$w$    & 0 & 1   & 25 & -3 & 0 & 0 & 0 \\ 
	$\rho$ & 1 & 0.2 & 25 & -4 & 4 & 7 & 14 \\
	$T$    & 1 & 0.2 & 25 & -5 & 5 & 6 & 15 \\
	\hline \hline
\end{tabular}
\caption{Parameters used for the 3D manufactured solution.}
\label{table:ParametersManuf3D}
\end{table}

Using the method of manufactured solutions (MMS) our first set of
examples demonstrates that the proposed solvers achieve the expected
temporal order of convergence. According to the MMS strategy, an arbitrary solution $Q$ is
prescribed, and a source term is added to the right hand side of
equation~\eqref{eq:NavierStokesCurvilinear3D} in such a way
that the proposed solution actually satisfies the equation.  For this
set of examples we use the MMS solution
$$ Q_j(\xi,\eta,\zeta,t) = \alpha_j + \beta_j \sin( 2\pi\lambda_j t + \phi_{j,t} ) \sin( 2\pi\xi + \phi_{j,\xi} ) \sin( 2\pi\eta + \phi_{j,\eta} ) \sin( 2\pi\zeta + \phi_{j,\zeta} ) $$
where $Q_j$ is the $j$th component of the solution vector
and where the various $\alpha_j$, $\beta_j$, $\lambda_j$, $\phi_{j,\cdot}$ are
constants. The parameter values we use for the solution are
given in Table~\ref{table:ParametersManuf3D}. The test geometry
in this context is a ``wavy cube'' given by the equations
\begin{eqnarray}
	x(\xi,\eta,\zeta) &=& \xi + a \left( \sin( 2\pi n \eta ) + \sin( 2\pi n \zeta ) \right) \nonumber \\
	y(\xi,\eta,\zeta) &=& \eta + a \left( \sin( 2\pi n \xi ) + \sin( 2\pi n \zeta ) \right) \nonumber \\
	z(\xi,\eta,\zeta) &=& \zeta + a \left( \sin( 2\pi n \xi ) + \sin( 2\pi n \eta ) \right) \nonumber
\end{eqnarray}
with $a=0.015$, $n=2$, and with $0\leq \xi,\eta,\zeta\leq 1$, which is
illustrated in Figure~\ref{fig:WavyCube}. Only the velocity components
($\mathbf u=0$) and temperature $T$ are prescribed at the boundary
according to the manufactured solution; the boundary values of the
density are obtained from the solution process, as described in
Section~\ref{sec:BCImplementation}. The Reynolds number and Mach
number in these examples are taken to equal $\mathrm{Re}=10^3$ and
$\mathrm{Ma} = 0.85$. The second- through sixth-order convergence of
the methods is demonstrated in Figure~\ref{fig:ManufConv3DWavy}. Note
in particular that the manufactured $T$ and $\rho$ solutions are
time-dependent on the boundary of the domain, thus demonstrating, in
particular, that the proposed numerical boundary condition
implementation preserves the correct order of time-accuracy even under
time-varying boundary values.

\begin{figure}[!htb]
	\centering
	\includegraphics[width=0.8\textwidth]{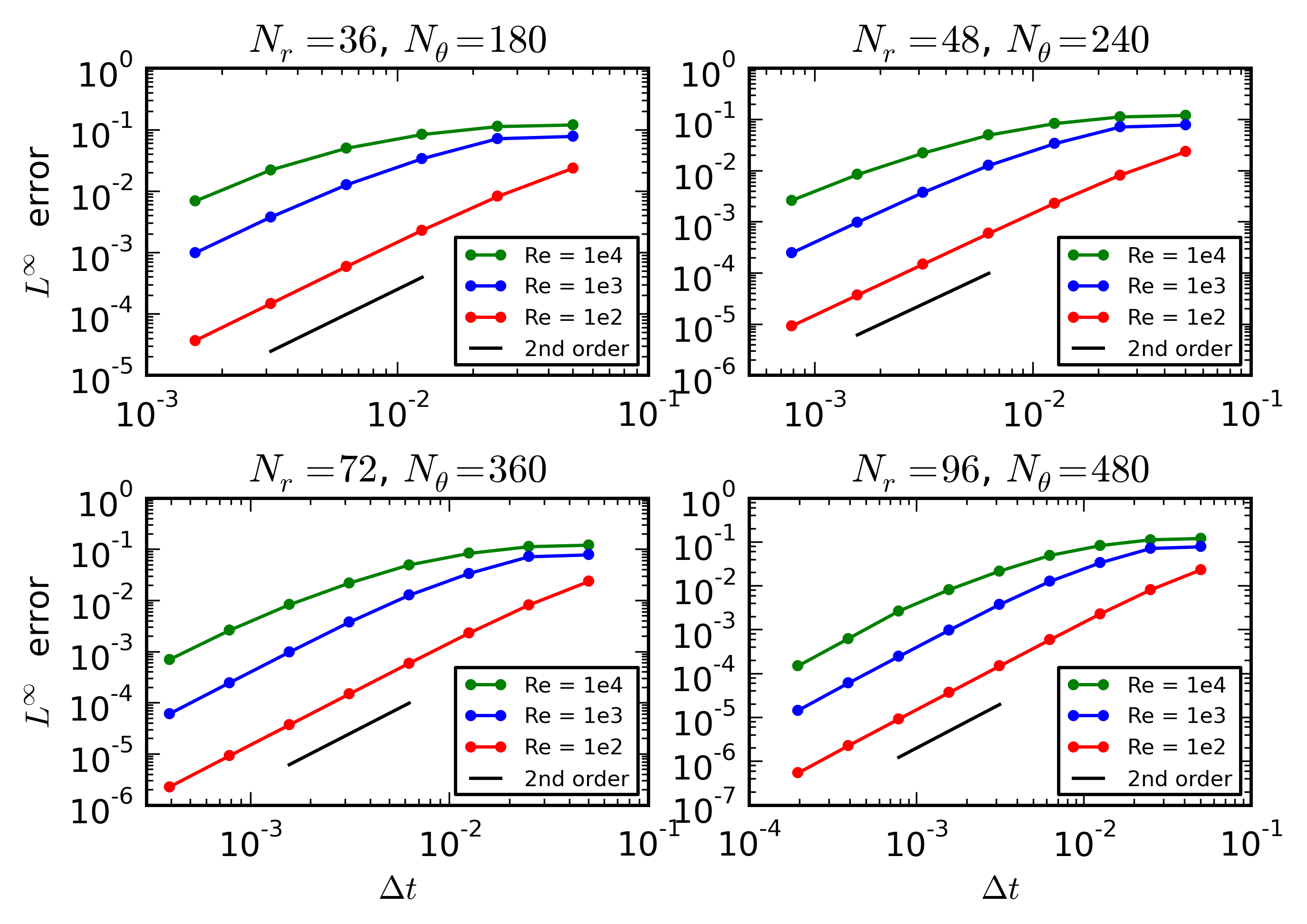}
	\caption{Convergence of the BDF2-ADI solver in an annulus for
          various mesh discretizations and Reynolds numbers.}
	\label{fig:AnnulusConvBDF2}
\end{figure}

\begin{figure}[!htb]
	\centering
	\includegraphics[width=0.8\textwidth]{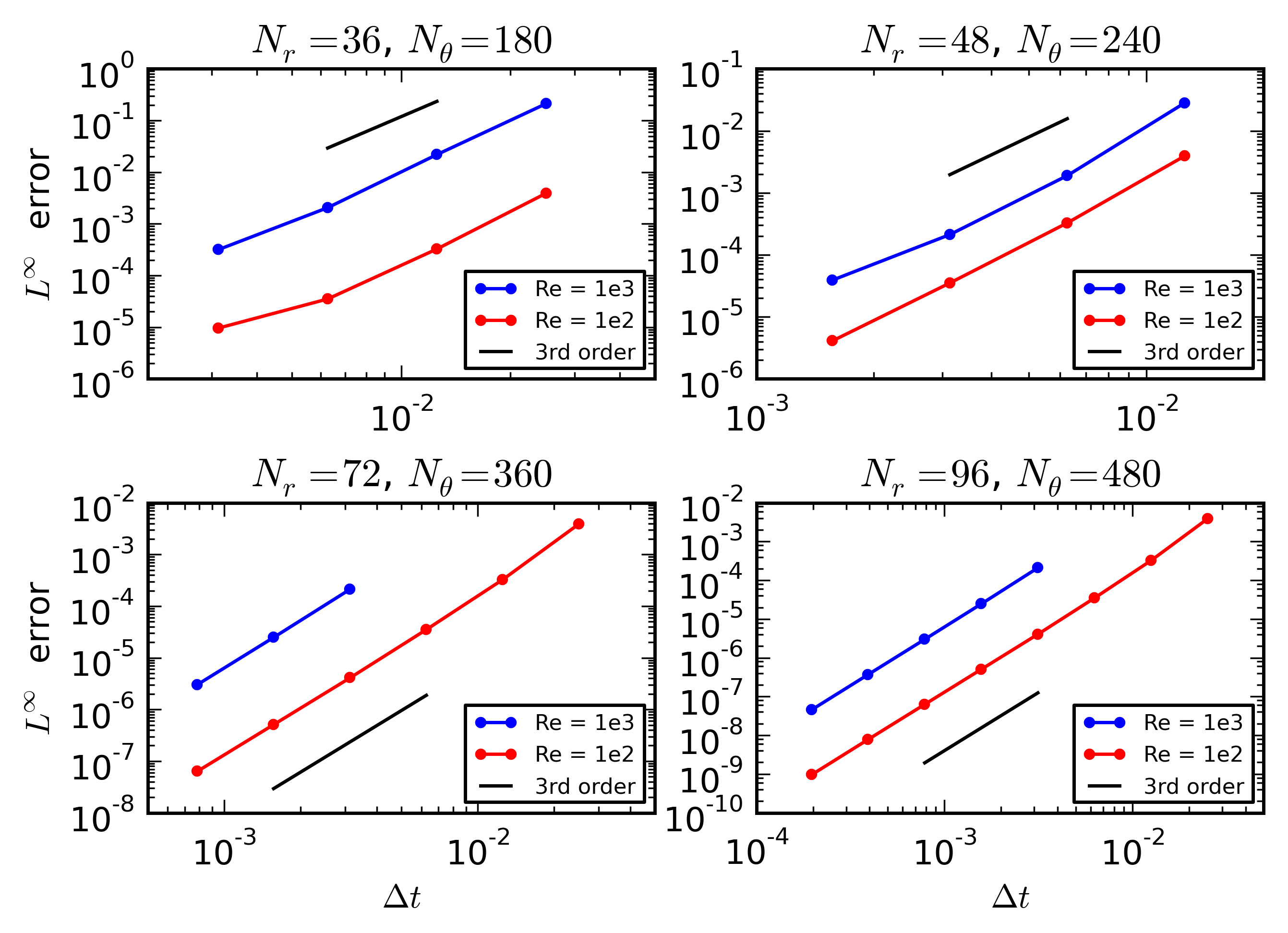}
	\caption{Convergence of the BDF3-ADI solver in an annulus with
          various mesh discretizations and Reynolds numbers.}
	\label{fig:AnnulusConvBDF3}
\end{figure}

Next, we demonstrate the convergence of the solver in two-dimensions with a
physical flow example at $\texttt{Ma}=0.8$ in an annulus with inner
radius $0.1$ and outer radius $0.5$ using Chebyshev collocation in the
radial direction and Fourier collocation in the azimuthal
direction. The flow starts with a zero initial condition for all
fields except temperature and density; the initial conditions for the
former are taken to equal $1.0$, while for the latter the initial
conditions are taken to equal the sum of the scalar $1.0$ plus two
Gaussian functions of the form
\begin{equation} \label{eq:GaussianEq}
a\exp \left( -\frac{(x-x_0)^2 + (y-y0)^2}{2 \sigma^2} \right)
\end{equation}
with parameters $a = 0.3$, $\sigma = 0.1$, $x_0 = -0.2$, $y_0 = 0.2$
and $a = -0.2$, $\sigma = 0.07$, $x_0 = 0.2$, $y_0 = 0$,
respectively. For time between $t=0$ and $t=0.5$, the rotation of the
inner cylinder is ramped up smoothly until it reaches a tangential
velocity of $1.0$. A temperature source term equal to ($\sin(2\pi t)$
times a Gaussian in space given by equation~\eqref{eq:GaussianEq} with
$a = 2.5$, $\sigma = 0.05$, $x_0 = -0.2$, $y_0 = -0.2$) is also
used. The convergence of the solver at time $t=1.0$ is estimated via
comparison with the solution obtained from a fine discretization ($N_r
= 108$, $N_\theta = 540$, $\Delta t = 0.1 \times 2^{-10}$). Figures
~\ref{fig:AnnulusConvBDF2} and ~\ref{fig:AnnulusConvBDF3} verify the
expected rates of convergence at various Reynolds numbers.

\begin{figure}[!htb]
	\centering
	\includegraphics[width=0.7\textwidth]{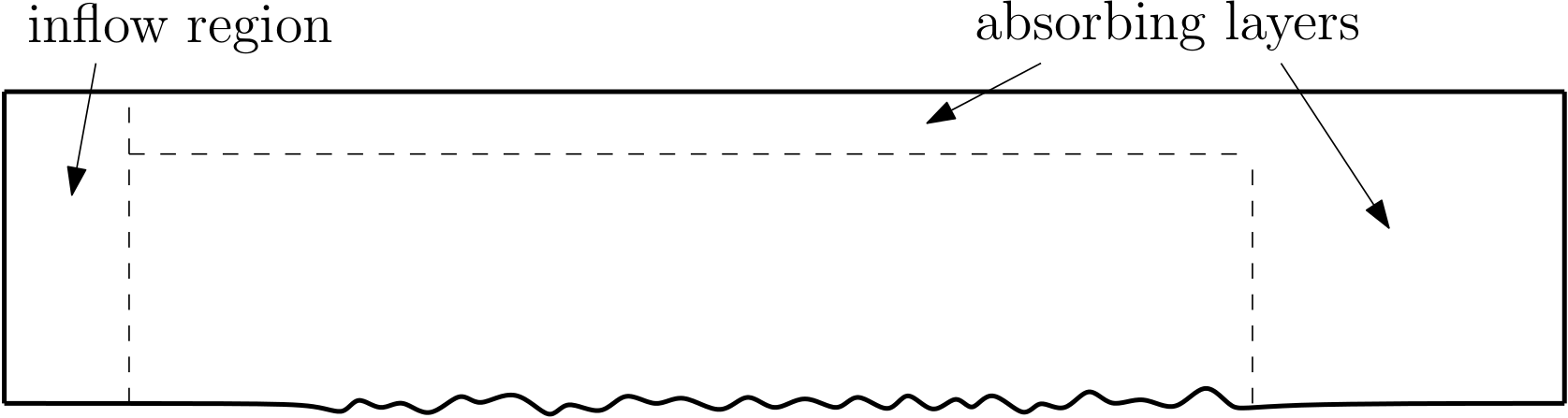}
	\caption{Schematic set-up of unsteady flow over a bumpy plate (not to scale).}
	\label{fig:BoundaryLayerSetup}
\end{figure}

\begin{figure}[!htb]
\centering
$\begin{array}{c}
	\includegraphics[width=0.8\textwidth]{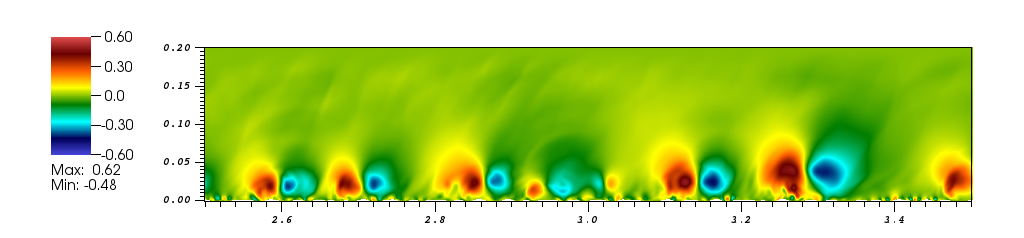}\\
	\includegraphics[width=0.8\textwidth]{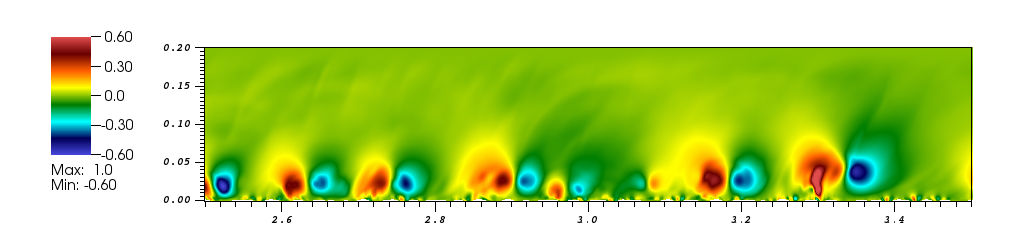}\\
	\includegraphics[width=0.8\textwidth]{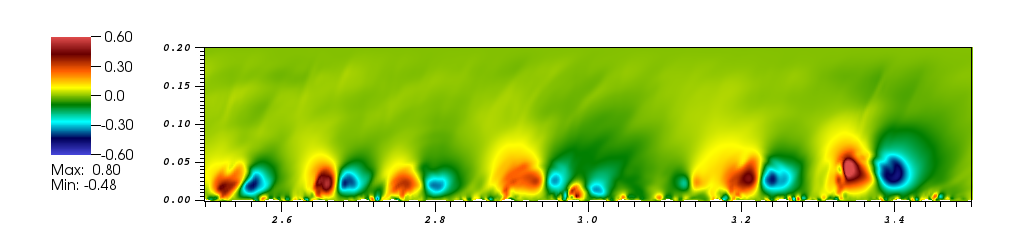}\\
	\includegraphics[width=0.8\textwidth]{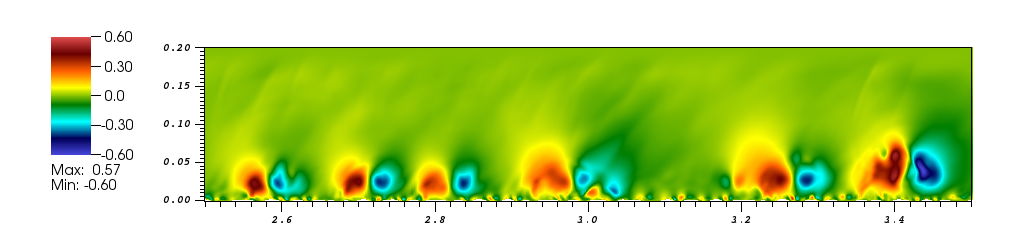}
\end{array}$
\caption{Evolution of the $y$-velocity component in two-dimensional
  boundary layer flow over a bumpy plate. From top to bottom, the
  solution times for the figures are $t = 9.76, \; 9.82, \; 9.88, \;
  9.94$.}
\label{fig:BumpyPlate}
\end{figure}

Next, a demonstration of boundary layer flow over a ``bumpy'' plate in
2D at high Reynolds number is presented. Here the domain is such that
the left and right edges of the domain lie on the lines $x= x_\ell=1$
and $x=x_r=4$, while the top and bottom boundaries lie on the curves
$y_t=0.2$ and 
$$ y_b(x) = \exp \left( -\left(\frac{x-2.5}{1.2}\right)^{12} \right)
\sum_{m=1}^4 a_m \sin( c_m x ), $$
respectively, where $a_1 = 5\times10^{-4}$, $a_2 =8\times10^{-4}$,
$a_3 = 6\times10^{-4}$, $a_4 = 4\times10^{-4}$, $c_1 = 300$, $c_2 =
207$, $c_3 = 161$ and $c_4 = 124$. The mesh in the interior of the
domain is generated by means of transfinite
interpolation~\cite{gordon_construction_1973}. A schematic
illustration of the set-up is provided in
Figure~\ref{fig:BoundaryLayerSetup}. A total of 1536 (resp. 96)
Chebyshev collocation points were used in the horizontal
(resp. vertical) direction .

To initialize the flow and impose boundary conditions we use the
asymptotic solution provided by the boundary layer equations for the
present compressible-flow
configuration~\cite[Ch. 7]{white_viscous_2006}. Here we provide a
brief overview in these regards; a more detailed discussion can be
found, e.g., in the aforementioned reference. For simplicity in the
solution of boundary-layer approximate equation we assume that the
viscosity and thermal conductivity are linear functions of temperature
($\mu(T) = \kappa(T) = T$), and that the Prandtl number equals unity:
$\mathrm{Pr}=1$. Using $x$ and $y$ coordinates tangent and normal to
the infinite planar boundary, the free-stream solution values as $y
\to \infty$ are assumed to equal $u_\infty = 1$, $v_\infty = 0$,
$T_\infty = 1$ and $\rho_\infty = 1$. The boundary layer equations are
obtained by transforming the steady ($Q_t=0$) two-dimensional
Navier-Stokes equations by means of the change of variables $y =
\delta Y$, where $\delta = \mathrm{Re}^{1/2}$ is the characteristic
length scale of the boundary layer. Furthermore, the solution
components are assumed to be perturbations of the free-stream values
of the form $u = u_\infty + \delta u_1$, $v = \delta v_1$, $T =
T_\infty + \delta T_1$, $\rho = \rho_\infty + \delta \rho_1$, which
leads to a set of equations for the inner solutions (terms with
subscript 1). Using the similarity variable $\bar{\eta} =
\bar{\eta}(x^{-1/2}Y)$ together with the Howarth
transformation~\cite{howarth_concerning_1948}, we obtain the following
simplified set of equations for $\bar{\eta}$, $u_1$, $v_1$, $\rho_1$,
and $T_1$ as functions of $x$ and $Y$:
\begin{gather*}
	\frac{\partial}{\partial Y}\bar{\eta} = \rho_1, \\
	\rho_1 \, u_1 = f'(\bar{\eta}), \\
	\rho_1 \, v_1 = \frac{1}{2}x^{-1/2} \left( \bar{\eta}\,f'(\bar{\eta}) - f(\bar{\eta}) \right), \\	
	T_1 = u_1 + T_\mathrm{wall} (1 - u_1) + \frac{1}{2}(\gamma-1)\mathrm{Ma}^2(u_1 - u_1^2), \\
	\rho_1 T_1 = 1,
\end{gather*}
where $T_\mathrm{wall}$ is the temperature at the wall and $f$ is the solution of the Blasius equation
\begin{gather*}
	f'''+ \frac{1}{2}f \, f' = 0, \\
	f(0) = f'(0) = 0, \\
	f'(\eta) \rightarrow 1 \; \mathrm{as} \; \eta \rightarrow \infty.
\end{gather*}
The similarity variable $\bar{\eta}$ is obtained by eliminating the
other unknowns and using a Newton-Kantorovich iterative
solver~\cite[App.  C]{boyd_chebyshev_2001} with initial guess computed
by standard fourth order Runge-Kutta. The remaining unknowns can then
be obtained explicitly from the above relations. The resulting
solution $Q_\mathrm{ref}$ of the boundary layer equations is used to
provide the initial condition and boundary conditions at inflow and,
as discussed in what follows, in the absorbing layers of the
computational domain as well. 

The boundary conditions for this example are no-slip conditions on the
bottom boundary ($\mathbf{u}_\mathrm{wall} = 0$ and $T_\mathrm{wall} =
1$), an absorbing layer of thickness 0.05 at the top of the domain,
another absorbing layer of thickness 0.5 is on the right, and inflow
conditions in a region of thickness 0.1 on the left; cf.
Figure~\ref{fig:BoundaryLayerSetup}. For each absorbing layer, a term
of the form $ \sigma(\xi,\eta) Q_\mathrm{ref} $ is added to the right
hand side of the PDE~\eqref{eq:NavierStokesCurvilinear3D} and $\sigma
I$ is added to the matrix $M^0$, where $I$ is the identity (cf. the
related, more elaborate absorbing-layer
method~\cite{appelo_high-order_2009}). The variable coefficient
$\sigma$ is given by
\begin{equation} \label{eq:Absorbing}
\sigma(\xi,\eta) = A \left( 1 - \psi \left( \frac{d(\xi,\eta)}{L} \right) \right)
\end{equation}
where $A$ is the absorption factor, $L$ is the width of the layer,
$d(\xi,\eta)$ is the distance to the boundary in question, and the function
$\psi$ is given by
\begin{equation} \label{eq:RampFunction}
	\psi(x) = \begin{cases}
			  	0 &, x \leq 0 \\
				1 &, x \geq 1 \\
				\left[1 + \exp \left( \frac{1}{x} - \frac{1}{1-x} \right)\right]^{-1} &, 0 < x < 1
			  \end{cases}
\end{equation}
For the top boundary we use $A = 50$, $L = 0.05$ and for the right
boundary we use $A = 20$, $L = 0.5$; these selections enforce adequate
damping in the absorbing layers. The boundary layer solution
$Q_\mathrm{ref}$ is prescribed in the inflow region for all
times. Figure \ref{fig:BumpyPlate} displays the $y$-velocity component
for various times, with $\mathrm{Re} = 10^6$, $\mathrm{Ma} = 0.85$,
and $\Delta t = 10^{-3}$. With this discretization ($\Delta
x_{\min}=3.1\times 10^{-6}$ and $\Delta y_{\min}=5.4\times 10^{-5}$),
an explicit solver would require a significantly smaller time step for
stability.

\begin{figure}[!htb]
	\centering
	\includegraphics[width=0.5\textwidth]{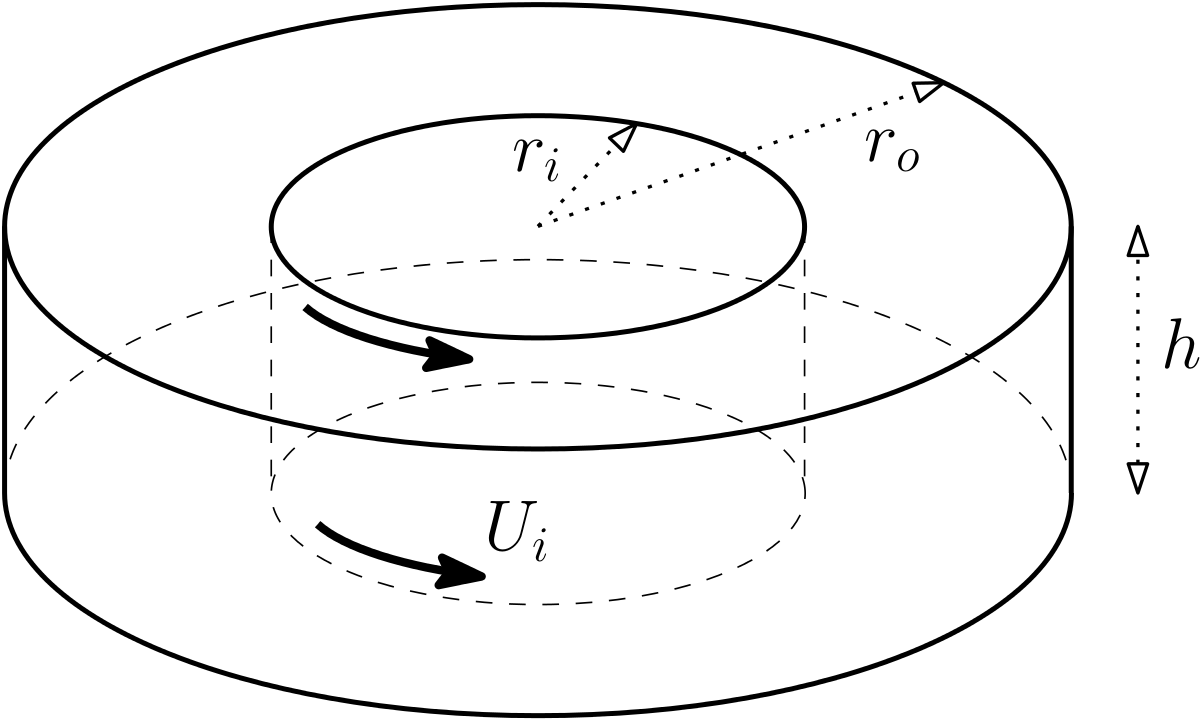}
	\caption{Geometry of the Taylor-Couette flow example. The
          fluid is confined in the region between the cylinders of
          radii $r_i$ and $r_o$ and between two planes separated by a
          distance $h$. The inner cylinder rotates with speed $U_i$
          while all other boundaries remain stationary.}
	\label{fig:TaylorCouetteSetup}
\end{figure}

\begin{figure}[!htb]
\centering
$\begin{array}{c|c|c}
	\includegraphics[width=0.3\textwidth]{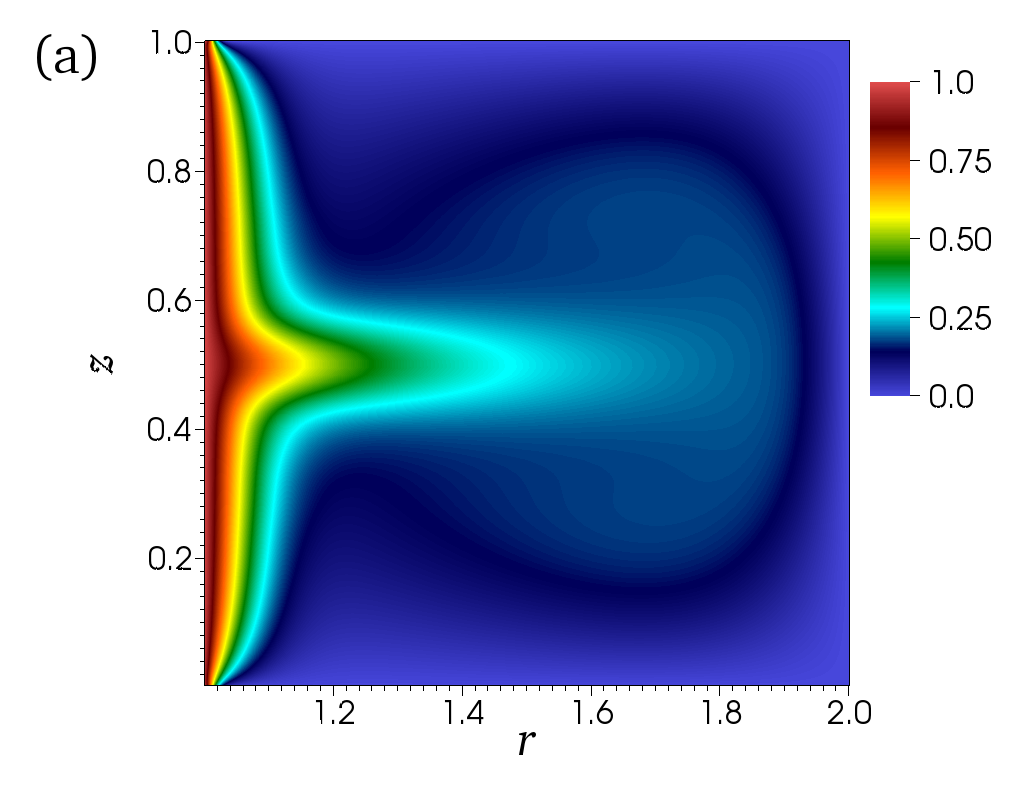}&
	\includegraphics[width=0.3\textwidth]{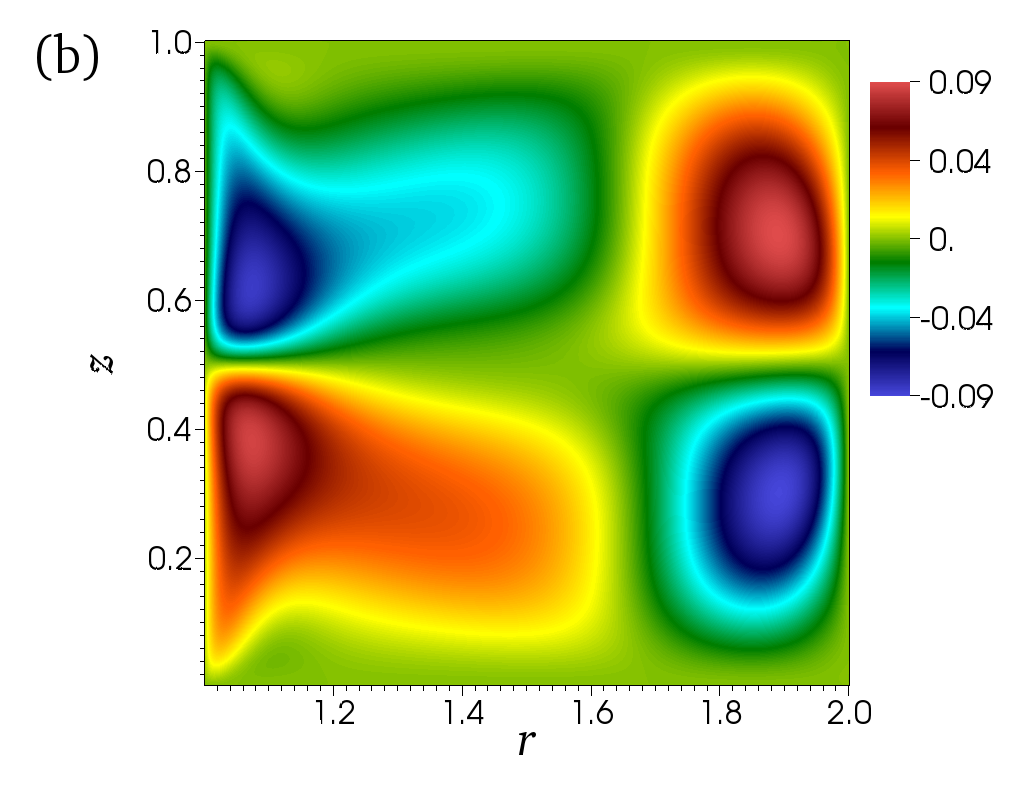}&
	\includegraphics[width=0.3\textwidth]{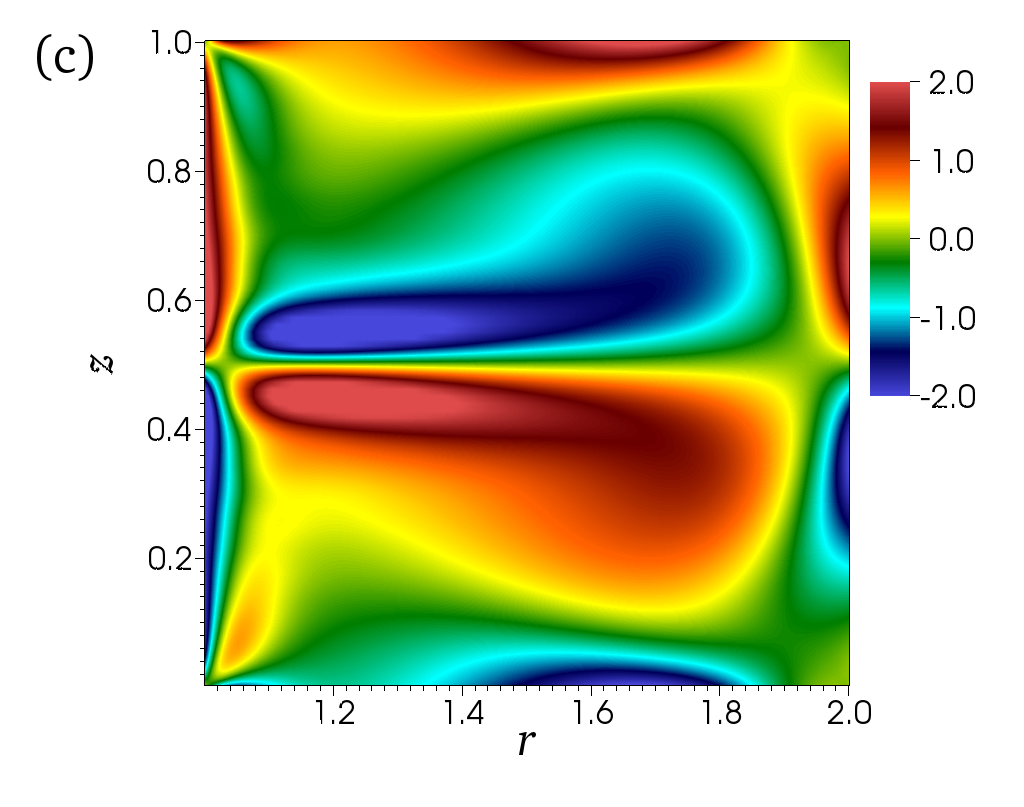} \\
	\includegraphics[width=0.3\textwidth]{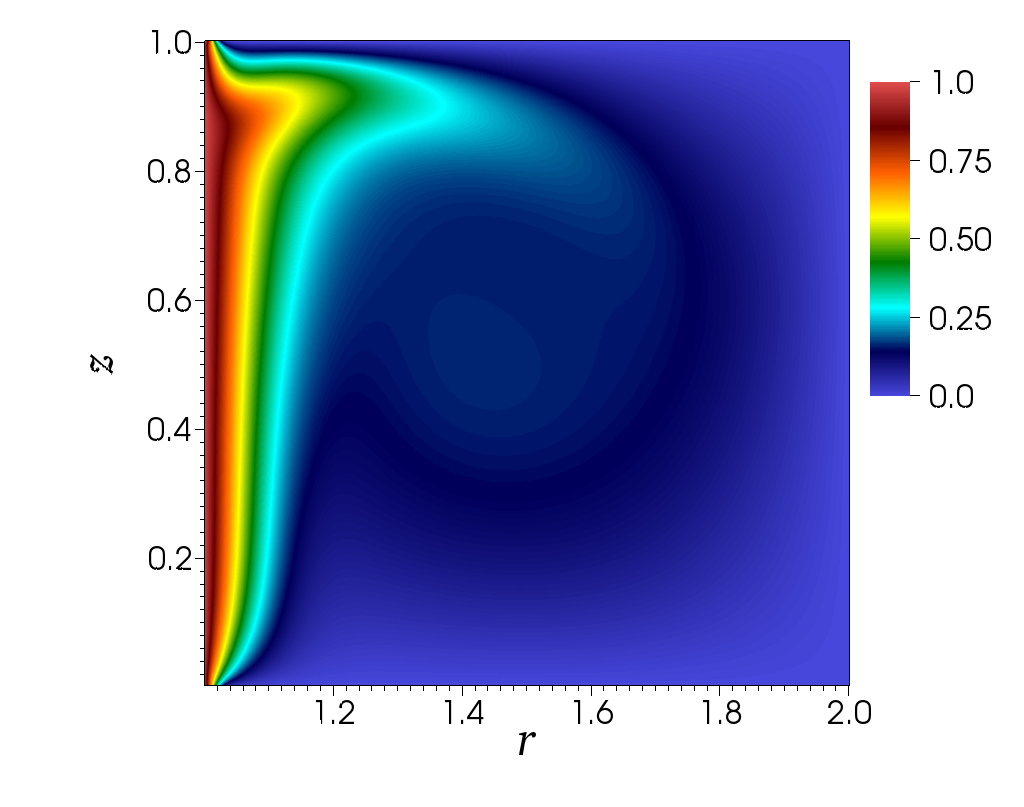}&
	\includegraphics[width=0.3\textwidth]{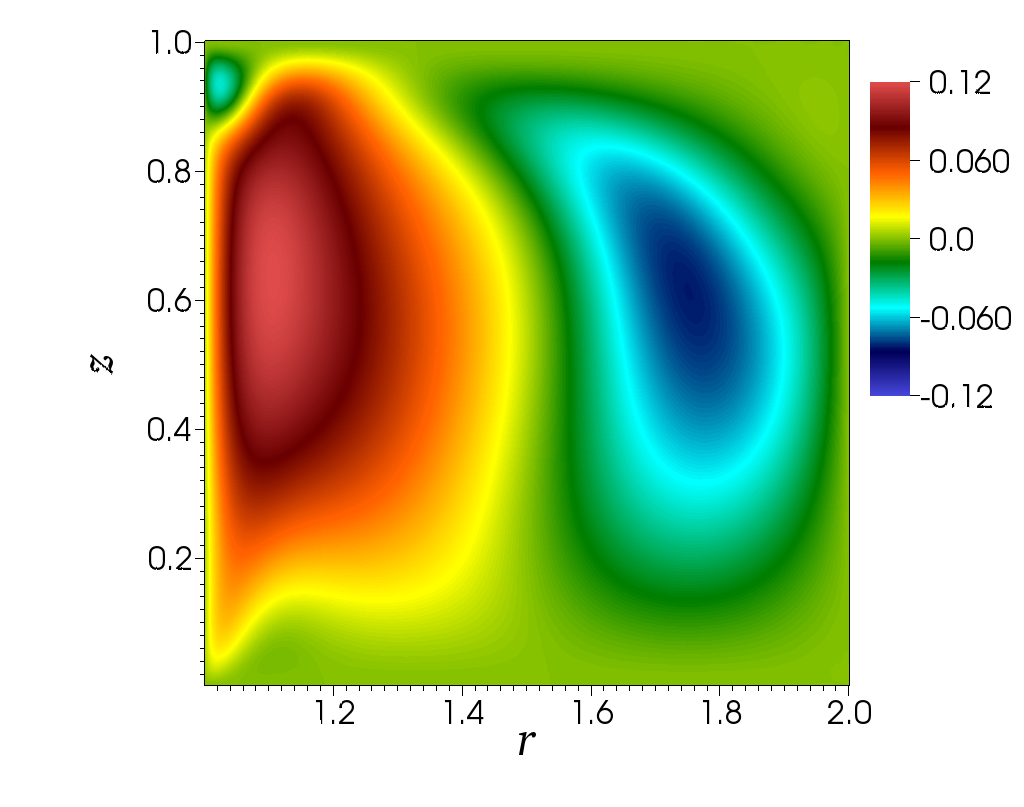}&
	\includegraphics[width=0.3\textwidth]{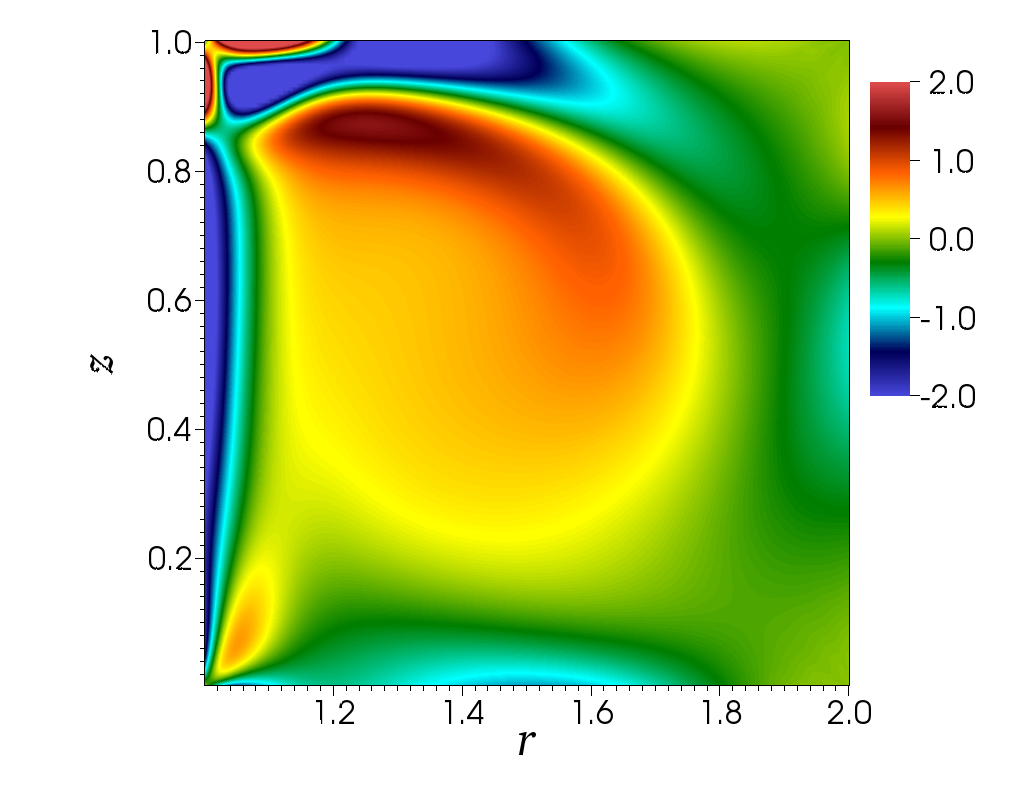}
\end{array}$
\caption{Profiles of the (a) azimuthal velocity, (b) vertical velocity
  and (c) azimuthal component of vorticity in small-aspect-ratio
  Taylor-Couette flow at $\mathrm{Ma} = 0.2$ and $\mathrm{Re} =
  700$. The top (resp. bottom) row displays profiles associated with
  the two-cell (resp. one-cell) stable mode.}
\label{fig:TCSmallAspect}
\end{figure}		

Our next example concerns three-dimensional simulations of Taylor-Couette flow, that
is to say, flow of a fluid between two concentric rotating cylinders,
as depicted in Figure~\ref{fig:TaylorCouetteSetup}. Most studies of
Taylor-Couette flow deal with incompressible fluids, but the dynamics
for subsonic compressible gases are similar, as shown in
\cite{manela_compressible_2007,kao_linear_1992}. The geometry is
defined by the inner radius $r_i$, outer radius $r_o$ and height
$h$. In what follows, we consider only the case where the inner
cylinder rotates and the outer cylinder together with the top and
bottom walls are stationary. In this case, the Reynolds number
$\mathrm{Re}$ is defined with respect to the velocity of the inner
cylinder.

The small aspect ratio regime ($\Gamma = \frac{h}{r_o - r_i} \approx
1$ or less) has been extensively studied both numerically and
experimentally (in the incompressible case)---in part because of the
property that two or more stable flows can exist for the same values
of the system parameters $\Gamma$, $\mathrm{Re}$, $r_o$,
$r_i$~\cite{pfister_bifurcation_1988,marques_onset_2006}. The
incompressible solution is
reported~\cite{pfister_bifurcation_1988,marques_onset_2006} to behave
as follows: For $\Gamma = 1$ and radius ratio $r_i/r_o = 0.5$, there
is only one stable flow at small $\mathrm{Re}$, which is characterized
by two axisymmetric toroidal vortices, one on top of the other,
similar to those depicted in the upper image in Figure
~\ref{fig:TCSmallAspect}~(c). At about $\mathrm{Re} \approx 133$, this
mode becomes unstable. The stable mode is then characterized by a
single large axisymmetric toroidal vortex in the center and a smaller
one in the inner upper corner, similar to those shown in the lower
image in Figure~\ref{fig:TCSmallAspect}~(c). Both modes are stable in
the range $603 \lesssim \mathrm{Re} \lesssim 786$.

Starting from the same initial condition (zero velocity, density and
temperature equal to $1.0$) and ending at the same final inner
cylinder rotational velocity with $\mathrm{Re}=700$ and $\mathrm{Ma} =
0.2$, our simulations produce both stable modes. To produce the
first mode, the inner cylinder velocity was varied as a function of
time according to
$$ U_i(t) = \psi \left( \frac{t}{160} \right) $$
where $\psi$ is defined in~\eqref{eq:RampFunction}. To produce the
second mode, in turn, the cylinder velocity was varied according to
the relation
$$ U_i(t) = 0.4 \psi \left( \frac{t}{10} \right) + 0.6 \psi \left( \frac{t-150}{150} \right).$$
The spatial discretization used a total of 48 Chebyshev collocation
points in the radial and $z$ directions and 64 Fourier collocation
points in the azimuthal direction. No-slip isothermal ($T=1$) boundary
conditions were used on all walls, with the angular velocity at the
top and bottom boundaries prescribed as
$$ u_\theta(r,t) = \exp \left( -\left( 2 \sqrt{\mathrm{Re}}(r - r_i) \right)^2 \right) U_i(t). $$
The time discretization for both simulations was set at $\Delta t =
0.02$ and simulations were stopped at $t=400$. At $\mathrm{Ma} = 0.2$,
there is less than 0.5\% deviation in the density from the initial
condition $\rho = T = 1$ throughout the simulations.  The presence of
corners in the geometry undoubtedly gives rise to reductions in the
solution accuracy (cf. Remark~\ref{rem:Corners}); nevertheless,
Figure~\ref{fig:TCSmallAspect} shows both modes at $t=300$, which
compares well to the experimental and numerical results in the
literature for the incompressible
case~\cite{pfister_bifurcation_1988,marques_onset_2006}---as it should
given the low value of the Mach number considered.

\section{Conclusions \label{sec:Conclusions}}

This paper introduced an implicit solution strategy for the
compressible Navier-Stokes equations which enjoys high-order accuracy
in time and which runs at spatial FFT speeds per time-step; of course,
the proposed BDF-ADI strategy can also be used in conjunction with
other spectral or non-spectral spatial approximations (such as
finite-differences, Fourier-Continuation, etc.). As emphasized above
in this text, the algorithms presented in this paper are the first
ADI-based Navier-Stokes solvers for which second order or better
accuracy has been verified in practice under non-trivial
(non-periodic) boundary conditions. The numerical examples presented
in this contribution demonstrate the favorable qualities inherent in
the proposed algorithms in both space and time.

\paragraph{Acknowledgments} The authors gratefully acknowledge support
from the Air Force Office of Scientific Research and the National Science
Foundation. MC also thanks the National Physical Science Consortium for
their support of this effort.

\appendix

\section{Quasilinear-like matrix coefficients in Cartesian
  and curvilinear coordinates} \label{app:NSMatrices}

Let $a = \frac{1}{\mathrm{Re}} \frac{\mu'(T)}{\rho}$, $b =
\frac{\gamma(\gamma-1)\mathrm{Ma}^2}{\mathrm{Re}} \frac{\mu(T)}{\rho}$, $c =
\frac{\gamma}{\mathrm{Re}\mathrm{Pr}} \frac{\kappa'(T)}{\rho}$, $d =
\frac{1}{\gamma \mathrm{Ma}^2}$, and $e = \gamma - 1$. The coefficient matrices
for Navier-Stokes equations in quasilinear-like Cartesian
form~\eqref{eq:NavierStokesCartesian3D} are
$$ M^x = 
\begin{pmatrix}
	u - \frac{2}{3} a T_x &
	-\frac{1}{2} a T_y &
	-\frac{1}{2} a T_z &
	d - a \left( u_x - \frac{1}{3} \nabla \cdot \mathbf{u} \right) &
	d \frac{T}{\rho} \\[1.5ex]
	
	\frac{1}{3} a T_y &
	u - \frac{1}{2} a T_x &
	0 &
	-\frac{1}{2} a (v_x + u_y) &
	0 \\[1.5ex]
	
	\frac{1}{3} a T_z &
	0 &
	u - \frac{1}{2} a T_x &
	-\frac{1}{2} a (w_x + u_z) &
	0 \\[1.5ex]
	
	e T - b \left( 2 u_x - \frac{2}{3} \nabla \cdot \mathbf{u} \right) & 
	-b (v_x + u_y) &
	-b (w_x + u_z) &
	u - c T_x &
	0 \\[1.5ex]
	
	\rho & 0 & 0 & 0 & u 
	
\end{pmatrix}
$$
$$ M^y = 
\begin{pmatrix}
	v - \frac{1}{2} a T_y &
	\frac{1}{3} a T_x &
	0 &
	-\frac{1}{2} a (v_x + u_y) &
	0 \\[1.5ex]
	
	-\frac{1}{2} a T_x &
	v - \frac{2}{3} a T_y &
	-\frac{1}{2} a T_z &
	d - a \left( v_y - \frac{1}{3} \nabla \cdot \mathbf{u} \right) &
	d \frac{T}{\rho} \\[1.5ex]
	
	0 &
	\frac{1}{3} a T_z &
	v - \frac{1}{2} a T_y &
	-\frac{1}{2} a (w_y + v_z) &
	0 \\[1.5ex]
	
	-b (v_x + u_y) &
	e T - b \left( 2 v_y - \frac{2}{3} \nabla \cdot \mathbf{u} \right) & 
	-b (w_y + v_z) &
	v - c T_y &
	0 \\[1.5ex]
	
	0 & \rho & 0 & 0 & v 
	
\end{pmatrix}
$$
$$ M^z = 
\begin{pmatrix}
	w - \frac{1}{2} a T_z &
	0 &
	\frac{1}{3} a T_x &
	-\frac{1}{2} a (w_x + u_z) &
	0 \\[1.5ex]
	
	0 &
	w - \frac{1}{2} a T_z &
	\frac{1}{3} a T_y &
	-\frac{1}{2} a (w_y + v_z) &
	0 \\[1.5ex]
	
	-\frac{1}{2} a T_x &
	-\frac{1}{2} a T_y &
	w - \frac{2}{3} a T_z &
	d - a \left( w_z - \frac{1}{3} \nabla \cdot \mathbf{u} \right) &
	d \frac{T}{\rho} \\[1.5ex]
	
	-b (w_x + u_z) &
	-b (w_y + v_z) &
	e T - b \left( 2 w_z - \frac{2}{3} \nabla \cdot \mathbf{u} \right) & 
	w - c T_z &
	0 \\[1.5ex]
	
	0 & 0 & \rho & 0 & w 
	
\end{pmatrix}
$$
$$ M^{xx} = -\frac{1}{\mathrm{Re}\,\rho} \; \mathrm{diag} \left( \frac{4}{3}\mu, \; \mu, \; \mu, \; \frac{\gamma}{\mathrm{Pr}}\kappa, \; 0 \right) $$

$$ M^{yy} = -\frac{1}{\mathrm{Re}\,\rho} \; \mathrm{diag} \left( \mu, \; \frac{4}{3}\mu, \; \mu, \; \frac{\gamma}{\mathrm{Pr}}\kappa, \; 0 \right) $$
$$ M^{zz} = -\frac{1}{\mathrm{Re}\,\rho} \; \mathrm{diag} \left(  \mu, \; \mu, \;\frac{4}{3}\mu, \; \frac{\gamma}{\mathrm{Pr}}\kappa, \; 0 \right) $$
The matrices $M^{xy}$, $M^{xz}$ and $M^{yz}$ are zero except for two elements
each, which are
$$ M^{xy}_{u,v} = M^{xy}_{v,u} = M^{xz}_{u,w} = M^{xz}_{w,u} = M^{yz}_{v,w} = M^{yz}_{w,v} = -\frac{1}{3} \frac{1}{\mathrm{Re}} \frac{\mu}{\rho}, $$

Using the above definitions and the metric terms $\xi_x$, $\xi_y$, etc. the
coefficient matrices in general coordinates for use
in~\eqref{eq:NavierStokesCurvilinear3D} are computed as
$$ M^\xi = \xi_x M^x + \xi_y M^y + \xi_z M^z + \xi_{xx} M^{xx} + \xi_{yy} M^{yy} + \xi_{zz} M^{zz} + \xi_{xy} M^{xy} + \xi_{xz} M^{xz} + \xi_{yz} M^{yz} $$
$$ M^{\xi\xi} = \xi_x^2 M^{xx} + \xi_y^2 M^{yy} + \xi_z^2 M^{zz} + \xi_x \xi_y M^{xy} + \xi_x \xi_z M^{xz} + \xi_y \xi_z M^{yz} $$
and $M^\eta$, $M^{\eta\eta}$ ($M^\zeta$, $M^{\zeta\zeta}$) are obtained by
replacing $\xi$ with $\eta$ ($\zeta$) in the above equations. The mixed
derivative matrices are computed as
$$ M^{\xi\eta} = \xi_x \eta_x M^{xx} + \xi_y \eta_y M^{yy} + \xi_z \eta_z M^{zz} + (\xi_x \eta_y + \eta_x \xi_y) M^{xy} + (\xi_x \eta_z + \eta_x \xi_z) M^{xz} + (\xi_y \eta_z + \eta_y \xi_z) M^{yz} $$
$$ M^{\xi\zeta} = \xi_x \zeta_x M^{xx} + \xi_y \zeta_y M^{yy} + \xi_z \zeta_z M^{zz} + (\xi_x \zeta_y + \zeta_x \xi_y) M^{xy} + (\xi_x \zeta_z + \zeta_x \xi_z) M^{xz} + (\xi_y \zeta_z + \zeta_y \xi_z) M^{yz} $$
$$ M^{\eta\zeta} = \eta_x \zeta_x M^{xx} + \eta_y \zeta_y M^{yy} + \eta_z \zeta_z M^{zz} + (\eta_x \zeta_y + \zeta_x \eta_y) M^{xy} + (\eta_x \zeta_z + \zeta_x \eta_z) M^{xz} + (\eta_y \zeta_z + \zeta_y \eta_z) M^{yz}. $$

\bibliography{ThesisBib}

\begin{thebibliography}{10}

\bibitem{albin_spectral_2011}
N.~Albin and O.~P. Bruno.
\newblock A spectral {FC} solver for the compressible
  {Navier}{\textendash}{Stokes} equations in general domains {I}: {Explicit}
  time-stepping.
\newblock {\em Journal of Computational Physics}, 230(16):6248--6270, July
  2011.

\bibitem{appelo_high-order_2009}
D.~Appel{\"o} and T.~Colonius.
\newblock A high-order super-grid-scale absorbing layer and its application to
  linear hyperbolic systems.
\newblock {\em Journal of Computational Physics}, 228(11):4200--4217, 2009.

\bibitem{beam_implicit_1978}
R.~M. Beam and R.~Warming.
\newblock An implicit factored scheme for the compressible {Navier}-{Stokes}
  equations.
\newblock {\em AIAA journal}, 16(4):393--402, 1978.

\bibitem{beam_implicit_1976}
R.~M. Beam and R.~F. Warming.
\newblock An implicit finite-difference algorithm for hyperbolic systems in
  conservation-law form.
\newblock {\em Journal of Computational Physics}, 22(1):87--110, Sept. 1976.

\bibitem{boyd_chebyshev_2001}
J.~P. Boyd.
\newblock {\em Chebyshev and {Fourier} spectral methods}.
\newblock Courier Dover Publications, 2001.

\bibitem{briley_structure_1980}
W.~R. Briley and H.~McDonald.
\newblock On the structure and use of linearized block implicit schemes.
\newblock {\em Journal of Computational Physics}, 34(1):54--73, Jan. 1980.

\bibitem{brown_overture:_1999}
D.~L. Brown, W.~D. Henshaw, and D.~J. Quinlan.
\newblock Overture: {Object}-oriented tools for overset grid applications.
\newblock {\em AIAA paper No. 99}, 9130, 1999.

\bibitem{bruno_quasi-unconditional_2015}
O.~P. Bruno and M.~Cubillos.
\newblock On the quasi-unconditional stability of {BDF}-{ADI} solvers for the
  compressible {Navier}-{Stokes} equations.
\newblock 2015.

\bibitem{bruno_higher-order_2014}
O.~P. Bruno and E.~Jimenez.
\newblock Higher-order linear-time unconditionally stable alternating direction
  implicit methods for nonlinear convection-diffusion partial differential
  equation systems.
\newblock {\em Journal of Fluids Engineering}, 136(6):060904--060904, Apr.
  2014.

\bibitem{bruno_high-order_2010}
O.~P. Bruno and M.~Lyon.
\newblock High-order unconditionally stable {FC}-{AD} solvers for general
  smooth domains {I}. {Basic} elements.
\newblock {\em Journal of Computational Physics}, 229(6):2009--2033, Mar. 2010.

\bibitem{bruno_spatially_2014}
O.~P. Bruno and A.~Prieto.
\newblock Spatially dispersionless, unconditionally stable
  {FC}{\textendash}{AD} solvers for variable-coefficient {PDEs}.
\newblock {\em Journal of Scientific Computing}, 58(2):331--366, 2014.

\bibitem{canuto_spectral_2007}
C.~Canuto, M.~Y. Hussaini, A.~Quarteroni, and T.~A. Zang.
\newblock {\em Spectral methods: evolution to complex geometries and
  applications to fluid dynamics}.
\newblock Springer Science \& Business Media, June 2007.

\bibitem{cubillos_general-domain_2015}
M.~Cubillos.
\newblock {\em General-domain compressible {Navier}-{Stokes} solvers exhibiting
  quasi-unconditional stability and high-order accuracy in space and time}.
\newblock {PhD}, California Institute of Technology, Mar. 2015.

\bibitem{dahlquist_special_1963}
G.~G. Dahlquist.
\newblock A special stability problem for linear multistep methods.
\newblock {\em BIT Numerical Mathematics}, 3(1):27--43, Mar. 1963.

\bibitem{douglas_two_1963}
J.~Douglas and J.~E. Gunn.
\newblock Two high-order correct difference analogues for the equation of
  multidimensional heat flow.
\newblock {\em Mathematics of Computation}, 17(81):71--80, 1963.

\bibitem{douglas_alternating_1962}
J.~Douglas, Jr. and J.~E. Gunn.
\newblock Alternating direction methods for parabolic systems in {M} space
  variables.
\newblock {\em J. ACM}, 9(4):450--456, Oct. 1962.

\bibitem{douglas_jr._general_1964}
J.~Douglas~Jr. and J.~E. Gunn.
\newblock A general formulation of alternating direction methods.
\newblock {\em Numerische Mathematik}, 6(1):428--453, Dec. 1964.

\bibitem{ekaterinaris_implicit_1999}
J.~A. Ekaterinaris.
\newblock Implicit, high-resolution, compact schemes for gas dynamics and
  aeroacoustics.
\newblock {\em Journal of Computational Physics}, 156(2):272--299, Dec. 1999.

\bibitem{garnier_large_2009}
E.~Garnier, N.~Adams, and P.~Sagaut.
\newblock {\em Large eddy simulation for compressible flows}.
\newblock Springer, 2009.

\bibitem{golub_matrix_2012}
G.~H. Golub and C.~F. Van~Loan.
\newblock {\em Matrix computations}, volume~3.
\newblock JHU Press, 2012.

\bibitem{gordnier_high_2009}
R.~E. Gordnier.
\newblock High fidelity computational simulation of a membrane wing airfoil.
\newblock {\em Journal of Fluids and Structures}, 25(5):897--917, July 2009.

\bibitem{gordon_construction_1973}
W.~J. Gordon and C.~A. Hall.
\newblock Construction of curvilinear co-ordinate systems and applications to
  mesh generation.
\newblock {\em International Journal for Numerical Methods in Engineering},
  7(4):461--477, 1973.

\bibitem{gottlieb_gibbs_1997}
D.~Gottlieb and C.-W. Shu.
\newblock On the {Gibbs} phenomenon and its resolution.
\newblock {\em SIAM Review}, 39(4):644--668, 1997.

\bibitem{hoffmann_computational_2000}
K.~A. Hoffmann and S.~T. Chiang.
\newblock Computational fluid dynamics, vol. 2.
\newblock {\em Engineering Education System, Wichita, Kansas}, pages 21--46,
  2000.

\bibitem{howarth_concerning_1948}
L.~Howarth.
\newblock Concerning the effect of compressibility on laminar boundary layers
  and their separation.
\newblock {\em Proceedings of the Royal Society of London. Series A.
  Mathematical and Physical Sciences}, 194(1036):16--42, July 1948.

\bibitem{kao_linear_1992}
K.-H. Kao and C.-Y. Chow.
\newblock Linear stability of compressible {Taylor}{\textendash}{Couette} flow.
\newblock {\em Physics of Fluids A: Fluid Dynamics (1989-1993)}, 4(5):984--996,
  May 1992.

\bibitem{kawai_large-eddy_2010}
S.~Kawai and S.~K. Lele.
\newblock Large-eddy simulation of jet mixing in supersonic crossflows.
\newblock {\em AIAA Journal}, 48(9):2063--2083, 2010.

\bibitem{kopriva_implementing_2009}
D.~A. Kopriva.
\newblock {\em Implementing spectral methods for partial differential
  equations: {Algorithms} for scientists and engineers}.
\newblock Springer Science \& Business Media, 2009.

\bibitem{lambert_numerical_1991}
J.~D. Lambert.
\newblock {\em Numerical methods for ordinary differential systems: the initial
  value problem}.
\newblock John Wiley \& Sons, Inc., 1991.

\bibitem{leveque_finite_2007}
R.~LeVeque.
\newblock {\em Finite difference methods for ordinary and partial differential
  equations}.
\newblock Society for Industrial and Applied Mathematics, Jan. 2007.

\bibitem{leveque_intermediate_1985}
R.~J. Leveque.
\newblock Intermediate boundary conditions for {LOD}, {ADI} and approximate
  factorization methods.
\newblock Technical {Report} NASA-CR-172591, ICASE-85-21, NASA, Mar. 1985.

\bibitem{lyon_high-order_2010}
M.~Lyon and O.~P. Bruno.
\newblock High-order unconditionally stable {FC}-{AD} solvers for general
  smooth domains {II}. {Elliptic}, parabolic and hyperbolic {PDEs}; theoretical
  considerations.
\newblock {\em Journal of Computational Physics}, 229(9):3358--3381, May 2010.

\bibitem{manela_compressible_2007}
A.~Manela and I.~Frankel.
\newblock On the compressible {Taylor}{\textendash}{Couette} problem.
\newblock {\em Journal of Fluid Mechanics}, 588:59--74, 2007.

\bibitem{marques_onset_2006}
F.~Marques and J.~M. Lopez.
\newblock Onset of three-dimensional unsteady states in small-aspect-ratio
  {Taylor}{\textendash}{Couette} flow.
\newblock {\em Journal of Fluid Mechanics}, 561:255--277, 2006.

\bibitem{peaceman_numerical_1955}
D.~W. Peaceman and H.~H. Rachford, Jr.
\newblock The numerical solution of parabolic and elliptic differential
  equations.
\newblock {\em Journal of the Society for Industrial and Applied Mathematics},
  3(1):28--41, Mar. 1955.

\bibitem{pfister_bifurcation_1988}
G.~Pfister, H.~Schmidt, K.~A. Cliffe, and T.~Mullin.
\newblock Bifurcation phenomena in {Taylor}-{Couette} flow in a very short
  annulus.
\newblock {\em Journal of Fluid Mechanics}, 191:1--18, 1988.

\bibitem{rizzetta_high-order_2008}
D.~P. Rizzetta, M.~R. Visbal, and P.~E. Morgan.
\newblock A high-order compact finite-difference scheme for large-eddy
  simulation of active flow control.
\newblock {\em Progress in Aerospace Sciences}, 44(6):397--426, Aug. 2008.

\bibitem{saad_gmres:_1986}
Y.~Saad and M.~H. Schultz.
\newblock {GMRES}: {A} generalized minimal residual algorithm for solving
  nonsymmetric linear systems.
\newblock {\em SIAM Journal on Scientific and Statistical Computing},
  7(3):856--869, 1986.

\bibitem{thomas_navier-stokes_1990}
P.~D. Thomas and K.~L. Neier.
\newblock Navier-{Stokes} simulation of three-dimensional hypersonic
  equilibrium flows with ablation.
\newblock {\em Journal of Spacecraft and Rockets}, 27(2):143--149, 1990.

\bibitem{uzun_simulation_2009}
A.~Uzun and M.~Y. Hussaini.
\newblock Simulation of noise generation in the near-nozzle region of a chevron
  nozzle jet.
\newblock {\em AIAA Journal}, 47(8):1793--1810, 2009.

\bibitem{visbal_high-order-accurate_1999}
M.~R. Visbal and D.~V. Gaitonde.
\newblock High-order-accurate methods for complex unsteady subsonic flows.
\newblock {\em AIAA Journal}, 37(10):1231--1239, 1999.

\bibitem{white_viscous_2006}
F.~M. White and I.~Corfield.
\newblock {\em Viscous fluid flow}, volume~3.
\newblock McGraw-Hill New York, 2006.

\end{thebibliography}

\end{document}